\documentclass[twocolumn,english,showpacs,preprintnumbers,amsmath,amssymb,floatfix]{revtex4-1}

\usepackage[T1]{fontenc}
\usepackage[latin9]{inputenc}
\usepackage{color}
\usepackage{array}
\usepackage{amstext}
\usepackage{graphicx}
\usepackage{esint}
\usepackage{rotating}
\usepackage{appendix}
\usepackage{float}

\usepackage[font=small,labelfont=bf]{caption}
\usepackage{xcolor}
\usepackage[colorlinks = true,
linkcolor = magenta,
urlcolor  = blue,
citecolor = red,
anchorcolor = blue]{hyperref}

\makeatletter

\providecommand{\tabularnewline}{\\}

\@ifundefined{textcolor}{}
{%
 \definecolor{BLACK}{gray}{0}
 \definecolor{WHITE}{gray}{1}
 \definecolor{RED}{rgb}{1,0,0}
 \definecolor{GREEN}{rgb}{0,1,0}
 \definecolor{BLUE}{rgb}{0,0,1}
 \definecolor{CYAN}{cmyk}{1,0,0,0}
 \definecolor{MAGENTA}{cmyk}{0,1,0,0}
 \definecolor{YELLOW}{cmyk}{0,0,1,0}
 }

\@ifundefined{definecolor}
 {\usepackage{color}}{}
\@ifundefined{definecolor}
 {\usepackage{color}}{}
\makeatother
\usepackage{babel}

\begin{document}

\title{ Study of spin-dependent structure functions of $^3{\rm He}$ and $^3{\rm H}$ at NNLO approximation and corresponding nuclear corrections }

\author{Hamzeh Khanpour$^{1,2}$}
\email{Hamzeh.Khanpour@mail.ipm.ir}

\author{S. Taheri Monfared$^{2}$}
\email{Sara.Taheri@ipm.ir}

\author{S. Atashbar Tehrani$^{3}$}
\email{Atashbart@gmail.com}

\affiliation {
	$^{(1)}$Department of Physics, University of Science and Technology of Mazandaran, P.O.Box 48518-78195, Behshahr, Iran  \\
	$^{(2)}$School of Particles and Accelerators, Institute for Research in Fundamental Sciences (IPM), P.O.Box 19395-5531, Tehran, Iran \\
	$^{(3)}$Independent researcher, P.O. Box 1149-8834413, Tehran, Iran  }

\date{\today}

%
%
\begin{abstract}\label{abstract}

We determine polarized parton distribution functions (PPDFs) and structure functions from recent experimental data of polarized deep inelastic scattering (DIS) on nuleons at next-to-next-to-leading order (NNLO) approximation in perturbative quantum chromodynamic (pQCD).
The nucleon polarized structure functions are computed using the Jacobi polynomial approach while target mass corrections (TMCs) are included in our fitting procedure.
Having extracted the polarized spin structure functions, we extend our study to describe $^3{\mathrm He}$ and $^3{\mathrm H}$ polarized structure functions, as well as the Bjorken sum rule.
We also explore the importance of the nuclear corrections on the polarized nuclear structure functions at small and large values of $x$.  Our results are compared with the recent available and high precision polarized $^3{\mathrm He}$ and $^3{\mathrm H}$ experimental data.

\end{abstract}

\maketitle
\tableofcontents{}

%
%
\section{Introduction}\label{Int}

Precise understanding of parton distribution functions (PDFs) will be a key ingredient in searches for new physics at the LHC through, for example, top-quark and Higgs-boson coupling measurements.
Consequently, reliable extraction of information on the unpolarized parton distribution functions  (PDFs)~\cite{Ball:2017nwa,Bourrely:2015kla,Harland-Lang:2014zoa,Martin:2009iq,Hou:2017khm}, polarized (PPDFs)~\cite{Shahri:2016uzl,Jimenez-Delgado:2014xza,Sato:2016tuz,Leader:2014uua,Nocera:2014gqa} and nuclear PDFs~\cite{Khanpour:2016pph,Eskola:2016oht,Kovarik:2015cma,Wang:2016mzo} from global QCD analyses of deep inelastic scattering (DIS) data, as well as many related studies~\cite{Bertone:2017tyb,Goharipour:2017rjl,Dahiya:2016wjf,Jimenez-Delgado:2013boa,Ball:2016spl,Goharipour:2017uic,Haider:2016zrk,Accardi:2016qay,Armesto:2015lrg,Frankfurt:2015cwa,Frankfurt:2016qca,MoosaviNejad:2016qdx,Nejad:2015fdh,Guzey:2016qwo,Khanpour:2017slc,Salajegheh:2015xoa,Kalantarians:2017mkj,Ethier:2017zbq,Kusina:2016fxy,Boroun:2015yea,Boroun:2014nia,Boroun:2014yea,Zarrin:2016kxf,AtashbarTehrani:2013qea,TaghaviShahri:2010zz,Phukan:2017lzp,Mottaghizadeh:2017vef}, provides deep understanding of the hadrons structure in term of their quarks and gluon constituents.

The determination of the longitudinal spin structure of the nucleon caused a huge growth of interest in polarized DIS experiments after
the surprising EMC~\cite{Ashman:1987hv,Ashman:1989ig} result that the quark spin contribution to the nucleon spin of $1/2$ might be significantly small.
In subsequent measurements by SMC~\cite{Adeva:1998vv}, it was confirmed that the quark spin contributes about one third of the spin of the nucleon. Many experiments have been conducted at SLAC, DESY and CERN to extract the nucleon spin-dependent structure functions $g_1(x,Q^2)$ and $g_2(x,Q^2)$. Various analyses of the world data of $A_1$ or $g_1$ based on next-to-leading order (NLO)~\cite{Jimenez-Delgado:2014xza,Sato:2016tuz,Leader:2014uua,Nocera:2014gqa} and next-to-next-to-leading order (NNLO)~\cite{Shahri:2016uzl,Khanpour:2017cha} calculations in perturbative quantum
chromodynamics (pQCD) have been carried out to extract the polarized parton densities along with the estimation of their uncertainties.

Previously in TKAA16~\cite{Shahri:2016uzl}, we carried out the first pQCD analysis using Jacobi polynomial approach at NNLO approximation based on only $g_1^{p,n,d}(x,Q^2)$ experimental data. In our latest study KTA-I17~\cite{Khanpour:2017cha}, we have extended our NNLO formalism by including target mass corrections (TMCs) and HT terms and enriched it by more data from $g_2^{p,n,d}(x,Q^2)$ observables. In this analysis also the Jacobi polynomials were implemented to determine the polarized parton distribution functions. This method has been applied to various QCD calculations~\cite{Shahri:2016uzl,Khanpour:2017cha,Ayala:2015epa,Leader:1997kw,MoosaviNejad:2016ebo,Khanpour:2016uxh}, containing the case of polarized and unpolarized PDF analyses.

In the absence of polarized charged current neutrino experiments, individual light quark sea densities cannot in principle be determined. The inclusive  polarized deep inelastic lepton-hadron reactions can only provide information about the $\Delta u+\Delta \bar{u}$, $\Delta d+\Delta \bar{d}$, and $\Delta s+\Delta \bar{s}$, along with the gluon.
For many years our group~\cite{Shahri:2016uzl,Khanpour:2017cha,Monfared:2014nta}, LSS~\cite{Leader:2002ni,Leader:2006xc}, BB10~\cite{Blumlein:2010rn}, and other people made simplifying assumptions about the sea quark densities and consequently were able to present results for the valence $u_v$ and $d_v$.
In the present study, we applied all polarized $g_1$ data including very recent COMPASS16 $g_1^p$ and $g_1^d$ data~\cite{Adolph:2015saz,Adolph:2016myg} to determine the sum of quark and anti-quark polarized PDFs $\Delta q(x)+\Delta \bar{q}(x)$. This method provides no information about the individual polarized quark and anti-quark distributions.
We focused only on $g_1$ experimental data due to their smaller uncertainties  compared with the $g_2$ measurements, indicating the lack of knowledge in the $g_2$ structure function.
Before one can precisely extract PPDF, it is important to take into account the target mass corrections arising from purely kinematic effects.

In addition to the scattering of polarized lepton beams from polarized nucleon, the polarized light nuclear targets provided the opportunity to probe the spin structure of the nucleon.
Among these are the SMC experiments at CERN~\cite{Adeva:1998vv} and the E143~\cite{Abe:1998wq} and E155~\cite{E155d} experiments at SLAC that used polarized deuterium. Meanwhile, the HERMES Collaboration at DESY~\cite{Ackerstaff:1997ws} and the E154 experiments at SLAC\cite{Abe:1997qk,Abe:1997cx} utilized polarized $^3{\mathrm He}$.
Recently both polarized $^3{\mathrm He}$ and $^3{\mathrm H}$ targets were used at E06-014 experiments at Jefferson Lab (JLAB) in Hall A which are the latest and most up-to-date data for the spin-dependent $g_1$ and $g_2$ structure functions of $^3{\mathrm He}$~\cite{Flay:2016wie}.
Hence, we step further from the ``structure of nucleon''  to  ``nuclei''  in terms of their parton constituents.
In order to study the polarized nuclear structure function, $g_{1,2}^A(x, Q^2)$, one needs to consider nuclear corrections. In this paper, we study the nuclear effects in inclusive scattering of polarized leptons from polarized $^3{\mathrm He}$ and $^3{\mathrm H}$ nuclei in the DIS region. We focus in particular on the kinematics at intermediate and large values of Bjorken $x$ where the major contributions come from the incoherent scattering.

This paper is organized as follows. In Sec.~\ref{Overview of data sets}, we provide an overview of data sets.
In Sec.~\ref{Proposed_methods}, we review our theoretical framework and summarize the basic formulas relevant for the analysis.
Section~\ref{uncertainty} contains the formalism used for computing the $\chi^2$ minimization and PPDF uncertainties.
We introduce the nuclear structure functions and corresponding nuclear corrections in Sec.~\ref{Nuclear-polarized-structure-functions}.
In Sec.~\ref{results} we discuss how well the predictions for the NNLO analysis and the inclusion of TMCs effects into NNLO polarized structure function analysis improves the precision of the extracted PPDFs  as well as nuclear structure functions.
Finally, in Sec.~\ref{Summary} we summarize our findings.

%
%
\section{Overview of data sets}\label{Overview of data sets}

The combined set of data was included in our NNLO QCD fit to the available $g_1^p$, $g_1^n$ and $g_1^d$ world data. For the proton data we use E143~\cite{Abe:1998wq},
HERMES98~\cite{HERM98}, HERMES06~\cite{HERMpd}, SMC~\cite{Adeva:1998vv}, EMC~\cite{Ashman:1987hv,Ashman:1989ig}, E155~\cite{E155p}, COMPASS10~\cite{COMP1} and COMPASS16~\cite{Adolph:2015saz}; for the neutron data we use E142~\cite{E142n},
E154~\cite{E154n}, HERMES98~\cite{HERM98}, HERMES06~\cite{Ackerstaff:1997ws}, JLAB03~\cite{JLABn2003}, JLAB04~\cite{JLABn2004} and JLAB05~\cite{JLABn2005}; and for the deuteron data we use E143~\cite{Abe:1998wq}, E155~\cite{E155d}, SMC~\cite{Adeva:1998vv}, HERMES06~\cite{HERMpd}, COMPASS05~\cite{COMP2005}, COMPASS06~\cite{COMP2006}, and COMPASS16~\cite{Adolph:2016myg}.
These data sets are summarized in Table~\ref{tab:DISdata}.
\begin{table*}[htb]
	\caption{Summary of published polarized DIS experimental data points with measured $x$ and $Q^2$ ranges and the number of data points, the $\chi^2$ for each given data set, and the fitted normalization shifts ${\cal{N}}_i$.} \label{tab:DISdata}
	\begin{ruledtabular}
		\begin{tabular}{l c c c c c c}
			\textbf{Experiment} & \textbf{Ref.} & \textbf{[$x_{\rm min}, x_{\rm max}$]}  & \textbf{Q$^2$ range {(}GeV$^2${)}}  & \textbf{Number of data points}  & $\chi^2$   &  \textbf{${\cal N}_n$}  \tabularnewline
			\hline\hline
			\textbf {E143(p)}   & \cite{Abe:1998wq}   & [0.031--0.749]   & 1.27--9.52 & 28 &21.424& 1.00034601\\
			\textbf{HERMES(p)} & \cite{HERM98}  & [0.028--0.66]    & 1.01--7.36 & 39 &75.369 & 1.00186515\\
			\textbf{SMC(p)}    & \cite{Adeva:1998vv}    & [0.005--0.480]   & 1.30--58.0 & 12 &10.803 & 0.99991146\\
			\textbf{EMC(p)}    & \cite{Ashman:1987hv}     & [0.015--0.466]   & 3.50--29.5 & 10 &3.328 & 1.00220752\\
			\textbf{E155}      & \cite{E155p}    & [0.015--0.750]   & 1.22--34.72 & 24 &35.170 & 1.024762208\\
			\textbf{HERMES06(p)} & \cite{HERMpd} & [0.026--0.731]   & 1.12--14.29 & 51 & 22.672 & 1.00018245\\
			\textbf{COMPASS10(p)} & \cite{COMP1} & [0.005--0.568]   & 1.10--62.10 & 15 &26.670 & 0.99301000\\
			\textbf{COMPASS16(p)} & \cite{Adolph:2015saz} & [0.0035--0.575]   & 1.03--96.1 & 54 &53.912 & 1.00019414\\
			\multicolumn{1}{c}{$\boldsymbol{g_1^p}$}  &       &  &  &  \textbf{233}  &  &  \\

			\textbf{E143(d)}  &\cite{Abe:1998wq}    & [0.031--0.749]   & 1.27--9.52    & 28  &38.159& 0.99916419\\
			\textbf{E155(d)}  &\cite{E155d}     & [0.015--0.750]   & 1.22--34.79   & 24&18.871 &0.99991576\\
			\textbf{SMC(d)}   &\cite{Adeva:1998vv}     & [0.005--0.479]   & 1.30--54.80   & 12 &18.375& 0.99998812\\
			\textbf{HERMES06(d)} & \cite{HERMpd}& []0.026--0.731]   & 1.12--14.29   & 51  &47.045& 1.00001347 \\
			\textbf{COMPASS05(d)}& \cite{COMP2005}& [0.0051--0.4740] & 1.18--47.5   & 11 &8.490 &0.99692499\\
			\textbf{COMPASS06(d)}& \cite{COMP2006}& [0.0046--0.566] & 1.10--55.3    & 15 & 12.874& 0.99991619\\
			\textbf{COMPASS16(d)} & \cite{Adolph:2016myg} & [0.0045--0.569]   & 1.03--74.1 & 43 &37.297  &1.00089129\\
			\multicolumn{1}{c}{ $\boldsymbol{g_1^d}$}    &    &  & & \textbf{184} &  &   \\

			\textbf{E142(n)}   &\cite{E142n}    & [0.035--0.466]   & 1.10--5.50    & 8 &7.466&0.99899932\\
			\textbf{HERMES(n)} &\cite{HERM98}   & [0.033--0.464]   & 1.22--5.25    & 9 &2.697 & 0.99995848\\
			\textbf{E154(n)}   &\cite{E154n}    & [0.017--0.564]   & 1.20--15.00   & 17&9.216 & 0.99961961\\
			\textbf{HERMES06(n)} &\cite{Ackerstaff:1997ws}  &  [0.026--0.731]  & 1.12--14.29   & 51 &17.974& 1.00001347\\
			\textbf{Jlab03(n)}&\cite{JLABn2003} & ]0.14--0.22]     & 1.09--1.46    & 4 &0.0469& 0.99981391\\
			\textbf{Jlab04(n)}&\cite{JLABn2004} & [0.33--0.60]      & 2.71--4.8     & 3 &3.651& 0.90000096\\
			\textbf{Jlab05(n)}&\cite{JLABn2005} & [0.19--0.20]     &1.13--1.34     & 2 & 1.674&1.02232189\\
			\multicolumn{1}{c}{$\boldsymbol{g_1^n}$}     &     &  & & \textbf{94}  &  &\\
			\hline\\
			\multicolumn{1}{c}{\textbf{ Total}}&\multicolumn{4}{c}{~~~~~~~~~~~~~~~~~~~~~~~~~~~~~~~~~~~~~~~~~~~~~~~~~~~~~~~~~~~~~~\textbf{511}}&\multicolumn{1}{c}{~~~~~~\textbf{473.195}}
			\\
		\end{tabular}
	\end{ruledtabular}
\end{table*}

\begin{figure*}[htb]
\vspace{1.0cm}
\includegraphics[clip,width=1.0\textwidth]{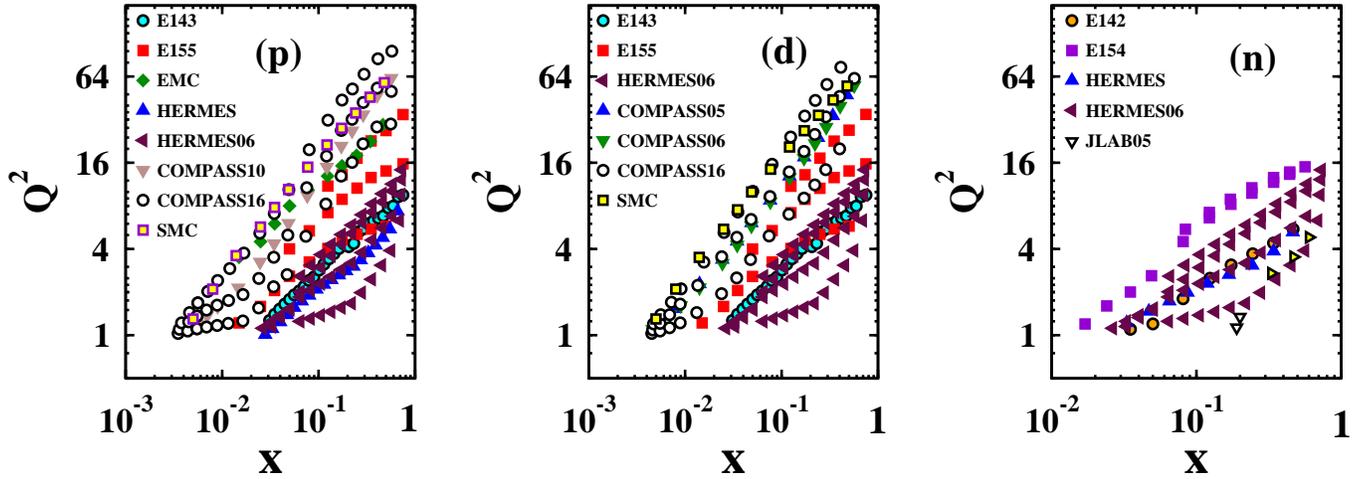}
\begin{center}
\caption{{Nominal coverage of the data sets used in the KTA-II17 analysis for proton, neutron, and deuteron observables. DIS data are presented on a logarithmic $x$ and $Q^2$ scales.  \label{fig:figxQ}}}
\end{center}
\end{figure*}

The $x$ and $Q^2$ nominal coverage of the data considered in our QCD fit is illustrated in Fig.~\ref{fig:figxQ}.
This plot nicely represents the kinematic coverage of $x$, and $Q^2$ of the proton, neutron and deuteron polarized structure functions.
%
Recent COMPASS results for protons and deuterons~\cite{Adolph:2015saz,Adolph:2016myg} at low $x$ ($x<0.03$) increase considerably the accuracy compared to the only available result in this region, SMC~\cite{Adeva:1998vv}.
Despite recent outstanding experimental efforts, the kinematical coverage of present polarized DIS observables is still rather limited.
Our data (511 experimental data points) cover the kinematic range
0.0035<$x$<0.75, 1\,(GeV$^2$)<$Q^2$<100 \, (GeV$^2$) and $W$>4 \, GeV.
	
This coverage leads to wider uncertainty for extracted polarized PDFs at small $x$.
Consequently, the polarized gluon distribution $\Delta G(x,Q^2)$ and strange distribution $(\Delta s +\Delta \bar{s})(x,Q^2)$ are still weakly constrained, especially for the case of the polarized gluon distribution.
Any conclusion on gluon treatment at  $x < 0.01$ relies on the behavior of low $x$ polarized DIS data, which is not accurately known up to now.
The statistical and systematic uncertainties are both taken into account. The systematic uncertainties are added quadratically.

%
%

\section{Theoretical analysis}\label{Proposed_methods}

%

%
\subsection{Leading-twist polarized DIS structure functions}

The calculations applied in this analysis are all performed at NNLO approximation. Correspondingly, the polarized PDFs  evolve from the input scale $Q_0^2$ using NNLO splitting functions~\cite{Moch:2014sna} and the NNLO hard scattering cross section expressions are considered.
To have a full leading-twist (LT) analysis we shall write
\begin{eqnarray}\label{g1:LT}
g_{1,2}(x,Q^2)_{\rm LT}& = & g_{1,2}(x,Q^2)_{\rm pQCD} + h_{1,2}^{\rm TMCs}(x,Q^2)/Q^2,  \nonumber \\
\end{eqnarray}
where $h_{1,2}^{\rm TMCs}(x,Q^2)$ is explained in Sec.~\ref{Target-mass-corrections}.
The NNLO spin-dependent proton structure functions, $g^{\rm p}_{1} (x, Q^2)_{\rm pQCD}$, can be written as a linear combination of polarized parton distribution functions $\Delta q$, $\Delta \bar{q}$ and $\Delta g$ as,
\begin{eqnarray}\label{eq:g1pxspace}
&& g_1^{\rm p} (x, Q^2)_{\rm pQCD} = \frac{1}{2} \sum_q e^2_q  \Delta q_{\rm NS}(x, Q^2) \otimes
\nonumber \\
&& \left(1+\frac{\alpha_s(Q^2)}{2 \pi}\Delta C^{(1)}_q + \left(\frac{\alpha_s(Q^2)}{2 \pi}\right)^2\Delta C^{(2)}_{\rm NS}\right)  \nonumber \\
&& + e^2_q (\Delta q + \Delta \bar{q})(x, Q^2)\otimes \nonumber \\
&& \left(1 + \frac{\alpha_s(Q^2)}{2 \pi}\Delta C^{(1)}_q + \left( \frac{\alpha_s(Q^2)}{2 \pi}\right)^2\Delta C^{(2)}_{s} \right) \nonumber \\
&& +\frac{2}{9}\left (\frac{\alpha_s(Q^2)}{2 \pi} \Delta C^{(1)}_g + \left( \frac{\alpha_s(Q^2)}{2 \pi}\right)^2\Delta C^{(2)}_g \right) \otimes \Delta g(x, Q^2), \nonumber \\
\end{eqnarray}
where the $\Delta C_q$ and $\Delta C_g$ are the spin-dependent quark and gluon hard scattering coefficient functions, calculable at NNLO approximation ~\cite{Lampe:1998eu,Zijlstra:1993sh}.
The symbol $\otimes$ represents the typical convolution in Bjorken $x$ space.  Considering the polarized proton structure function, we can apply isospin symmetry to achieve the neutron one. The deuteron structure function is related to that of the proton and neutron via
\begin{eqnarray}\label{eq:g1dxspace}
g_1^{\rm d}(x, Q^2)_{\rm LT} && = \frac{1}{2}\{g_{1}^{\rm p}(x, Q^2)_{\rm LT} + g_1^{\rm n}(x, Q^2)_{\rm LT}\}  \nonumber  \\
&& \times (1 - 1.5 w_D) \,,
\end{eqnarray}
where $w_D = 0.05 \pm 0.01$ is the probability to find the deuteron in a $D$-state~\cite{Lacombe:1981eg,Buck:1979ff,Zuilhof:1980ae}.

One can use the Wandzura and Wilczek (WW)~\cite{Wandzura:1977qf,Flay:2016wie} approximation for the leading-twist $g_2$ polarized structure function
\begin{eqnarray}\label{eq:WW}
&& g_2(x, Q^2)_{\rm pQCD}  =  g_2 ^{WW}(x, Q^2) =  \nonumber  \\
&& - g_1(x, Q^2)_{\rm pQCD} + \int_x^1\frac{dy}{y} g_1(y, Q^2)_{\rm pQCD}\,. 
\end{eqnarray}
Target mass corrections do not affect the WW relation if all powers in $(M^2/Q^2)$ are included~\cite{Wandzura:1977qf}.

%
%
\subsection{Target mass corrections}\label{Target-mass-corrections}

As illustrated in Fig. \ref{fig:figxQ}, in polarized DIS most of the small $x$ experimental data points are at low $Q^2$, which is one of the features of polarized DIS. In the unpolarized case we can cut the preasymptotic region data, while it is impossible to perform such a procedure for present data on spin-dependent structure functions without losing too much information.
To perform a reliable QCD fit including data at lower $Q^2$ values, target mass corrections cannot be ignored.
The standard approach to calculate TMCs  in the case of unpolarized DIS is the one based on the operator product  expansion (OPE) in QCD, first formulated by Georgi and Politzer~\cite{Georgi:1976ve}. This closed-form expression is generalized in Ref.~\cite{Blumlein:1998nv}.

We have performed an analysis to higher terms in the TMC expansion based on the method of Ref.~\cite{Sato:2016tuz}
and found that these terms do not change the agreement with $g_1$ data and the extracted polarized parton densities are insensitive to such a choice.
We claim that leading terms in the TMC expansion are reliable in the kinematical range of presently available $g_1$ data. In our previous study, KTA-I17~\cite{Khanpour:2017cha}, we presented the significant effect of considering TMCs and HT contributions while $g_2$ structure function data are included.

To consider the full LT approach Eq.~(\ref{g1:LT}) in our analysis, we applied the method suggested in Refs.~\cite{Piccione:1997zh,Dong:2006jm,Nachtmann:1973mr}. This method effectively depends on the LT term (for more details, see
our paper~\cite{Khanpour:2017cha}). For simplicity of notation, from now  to the end of the paper, we will drop the subscript ``LT'' denoting the leading twist.

%
%
\subsection{PDF parametrizations and conventions}

The method applied to reconstruct the $x$-dependent quantities from their Mellin moments is the Jacobi polynomial method (the same as our previous QCD analyses~ \cite{Shahri:2016uzl,Khanpour:2017cha}). The main difference, as  indicated in Sec. \ref{Int}, is that we consider a new input parametrization at the initial scale Q$_0^2$ = 1 GeV$^2$
for the sum of quark and antiquark polarized PDF instead of the valence and sea distributions, which are more general. 
We consider the general form of
\begin{eqnarray} \label{input_PDFs}
\nonumber
x(\Delta u + \Delta \bar{u})(x, Q^2_0) &=& \eta_{u_+} A_{u_+} x^{\alpha_{u_+}} (1-x)^{\beta_{u_+}}  \\ [2mm] \nonumber
&& \times (1 + \epsilon_{u_+}{\sqrt{x}} + \gamma_{u_+}x),  \\ [2mm]  \nonumber
x(\Delta d + \Delta \bar{d})(x, Q^2_0) &=& \eta_{d_+}A_{d_+}x^{\alpha_{d_+}} (1 - x)^{\beta_{d_+}}(1+\gamma_{d_+}x), \\ [2mm] \nonumber
x(\Delta s + \Delta \bar{s})(x, Q^2_0) &=& \eta_{s_+}A_{s_+}x^{\alpha_{s_+}} (1 - x)^{\beta_{s_+}}, \\ [2mm]\nonumber
x\Delta G(x, Q^2_0) & = & \eta_G A_Gx^{\alpha_G} (1 - x)^{\beta_G}\\ [2mm]
&& \times (1 + \epsilon_{G}{\sqrt{x}} + \gamma_{G}x).
\end{eqnarray}
Here, the notation $q_+=q+\bar{q}$ is applied for light quarks. The normalization factors, $A_{i}$, are fixed such that $\eta_i$ represent the first moments of the polarized distributions. As usual, the set of free parameters in Eq.~\ref{input_PDFs} is constrained by the well-known sum rules
\begin{eqnarray} \label{sumrule:1}
&&a_3 = g_A = \text{F + D} = 1.269 \pm 0.003, \nonumber \\
&&a_8 = \text{3F - D} = 0.585 \pm 0.025.
\end{eqnarray}
Here, $a_3$ and $a_8$ are nonsinglet combinations of the first moments of the polarized parton distributions corresponding to the axial charges for octet baryons~\cite{Bass:2009ed,Patrignani:2016xqp}. These parameters, F and D, are measured in hyperon and neutron $\beta$ decay and finally lead to the constraints
\begin{eqnarray}\label{axialvector:2}
a_3 &=& (\Delta u + \Delta \bar{u})(Q^2) - (\Delta d + \Delta \bar{d})(Q^2),  \nonumber \\
a_8 &=& (\Delta u + \Delta \bar{u})(Q^2) + (\Delta d + \Delta \bar{d})(Q^2)   \nonumber \\
&& -2 (\Delta s + \Delta \bar{s})(Q^2).
\end{eqnarray}
So considering  Eqs.~(\ref{sumrule:1})and~(\ref{axialvector:2}), the parameters $\eta_{u_+}$ and $\eta_{d_+}$ can be extracted versus $\Delta u + \Delta \bar{u}$, $\Delta d + \Delta \bar{d}$, and $\Delta s + \Delta \bar{s}$.
Here, we do not make any simplifying assumptions on the equality of the light sea quark distributions.

In our previous papers~\cite{Khanpour:2017cha,Shahri:2016uzl},
we have considered the Jacobi polynomial method to yield the structure functions from their Mellin moments in N space.

In the polynomial approach, one can easily expand the polarized structure functions in terms of the Jacobi polynomials $\Theta_{n}^{\alpha, \beta}(x)$ as follows,
\begin{equation}\label{xg1QCD}
	x\,g_1(x,Q^2)=x^{\beta}(1-x)^{\alpha}\,\sum_{n=0}^{N_{\rm max}} a_n(Q^2) \, \Theta_n^{\alpha,\beta}(x) \,.
\end{equation}
Here, $\Theta_n^{\alpha, \beta}(x) = \sum_{j=0}^n \, c_j^{(n)}(\alpha,\beta) \, x^j$ and $N_{\rm max}$ is the maximum order of the expansion.

The Jacobi polynomials $\Theta_n^{\alpha, \beta}(x)$ are a class of classical orthogonal polynomials. They are orthogonal with respect to the weight $x^{\beta} (1-x)^{\alpha}$ on the interval [0, 1].
In the polynomial approach, the $Q^2$ dependence of the $x \, g_{1}(x,Q^2)$ are codified in the Jacobi polynomial moments, $a_n(Q^2)$. Using the orthogonality relation, one can obtain this moment as
\begin{eqnarray}\label{aMoment}
	a_n(Q^2) &=& \int_0^1 dx\,x g_1(x,Q^2)\,\Theta_n^{\alpha,\beta}(x)   \nonumber \\
	&=& \sum_{j=0}^n\,c_j^{(n)}(\alpha,\beta)\,{\cal M} [xg_{1},\,j+2](Q^2)\,.
\end{eqnarray}
 The Mellin transform ${\cal {M}} [x g_1^{\tau 2}, \rm N]$ is defined via,
\begin{eqnarray}\label{Mellin}
	{\cal {M}} [xg_1,{\rm N}](Q^2) \equiv \int_{0}^{1}dx\,x^{\rm N-2}\,xg_1(x,Q^2)\,.
\end{eqnarray}
Applying the Jacobi polynomial expansion method, the polarized structure function $x g_1(x, Q^2)$ can be constructed as
\begin{eqnarray}\label{xg1Jacobi}
	x g_1(x,Q^2) &=& x^{\beta}(1-x)^{\alpha}\,\sum_{n=0}^{N_{max}}\,\Theta_n^{\alpha,\beta}(x)  \nonumber \\
	&\times& \sum_{j=0}^n\,c_j^{(n)}{(\alpha,\beta)}\,{\cal M}[x g_1, j+2] (Q^2) \,.
\end{eqnarray}
We also scrutinized the sensitivity of the Jacobi polynomials to its free parameters. These results are discussed in detail in Ref.~\cite{Khanpour:2017cha}. In our current analysis by setting the \{$N_{\rm max}$=7, $\alpha$=3, $\beta$=0.5\}, the optimal convergence of this expansion throughout the whole kinematic region covered by the polarized DIS data is achievable.

%
%
\section{$\chi^2$ minimization and uncertainties of physical predictions}\label{uncertainty}

The goodness of fit is traditionally determined by the effective global $\chi^2$ minimization algorithm that measures the quality of fit between theory and experiment.
$\chi_{\mathrm {global}}^2$ is defined by
\begin{eqnarray}\label{eq:chi2global}
&&\chi_{\mathrm {global}}^2 =  \sum_{i=1}^{N_n^{\mathrm {data}}} w_n \times \, \nonumber \\
&&\left[ \left( \frac{1 -{\cal N}_n }{\Delta{\cal N}_n}\right)^2 + \sum_{i=1}^{N_n^{\mathrm {data}}} \left(\frac{{\cal N}_n  \, g_{1, i}^{\mathrm {Exp}} - g_{1, i}^{\mathrm {Theory}} }{{\cal N}_n \, \Delta g_{1, i}^{\mathrm {Exp}}} \right)^2\right]\,,
\end{eqnarray}
where $n$ labels the number of different experiments and $w_n$ is a weight factor for the $n$th experiment. $g^{\mathrm {Exp}}$, $\Delta g^{\mathrm {Exp}}$, and $g^{\mathrm {Theory}}$ indicate the data value, the uncertainty, and the theory value for the data point $i$ of data set $n$, respectively. The ${\Delta{\cal N}_n}$ are the experimental normalization uncertainty quoted by the experiments. The  relative normalization factors,  ${\cal N}_n$, appear as free parameters in the fit.
These 22 normalization shifts are determined at the prefitting procedure along with the PDF free parameters and strong coupling constant using the CERN program library MINUIT~\cite{James:1994vla}. Afterwards, they are fixed at their best fitted values to further reduce the free parameters. Finally,
we minimize the above $\chi^2$ value with 16 free parameters including the strong coupling constant.

\begin{figure}[htb]
	\includegraphics[clip,width=0.49\textwidth]{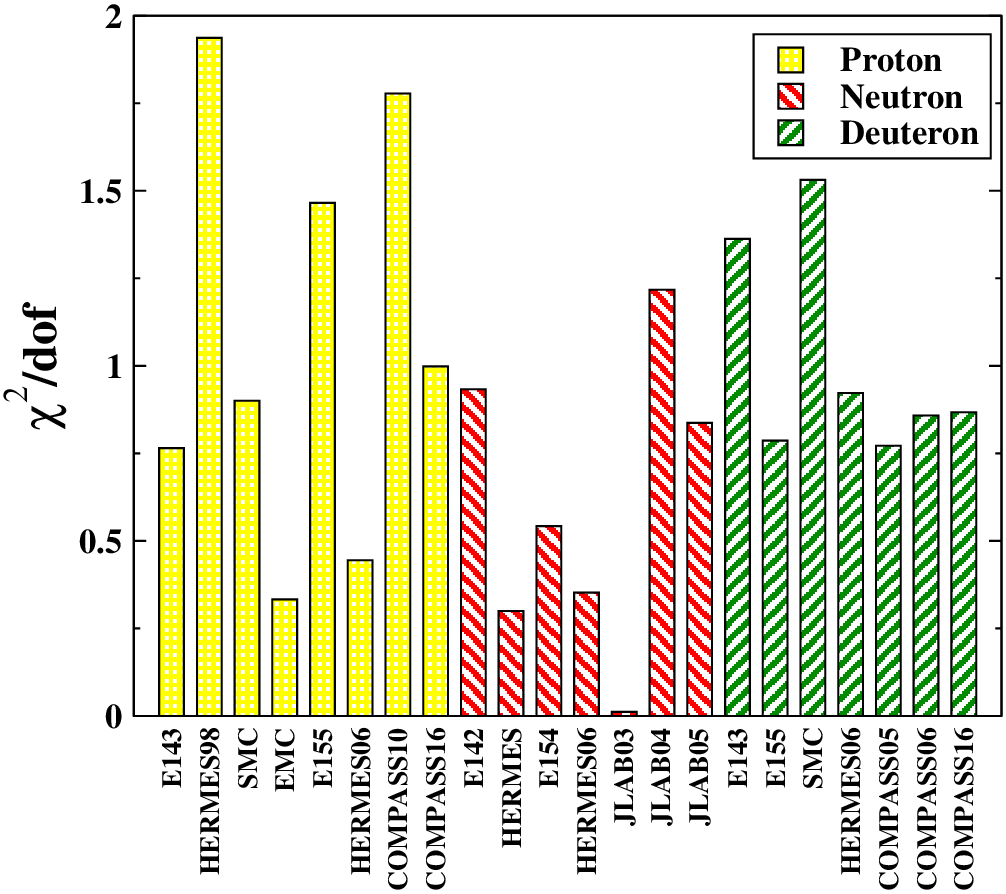}
	\begin{center}
		\caption{{Value of $\chi^2/dof$ for individual experiments per nucleon target used in the KTA-II17 \label{fig:chi2}}}
	\end{center}
\end{figure}

To visually evaluate the fit quality, in Fig.~\ref{fig:chi2} we plot the $\chi^2/dof$ for individual experiments per nucleon target. This allows us to check that the majority of experiments have a $(\chi^2/dof)\simeq1$. It means that most of the experiments satisfy this goodness-of-fit scale parameter.
The largest contribution to $\chi^2$ arises from the HERMES and COMPASS10 data for protons, and the SMC data for deuteron structure functions.  The smallest contribution comes from the
JLAB03 data for $xg_1^n$. From Fig.~\ref{fig:chi2},  one can conclude that the HERMES98 and COMPASS10 data for  $xg_1^p$ are difficult to be describe based on our fitting scenario. The motivation of considering them in our analysis originates mainly from our goal to have the most complete and up-to-date sets of data for the polarized structure functions.

Different methods to estimate the uncertainties of PDFs obtained from global  $\chi^2$ optimization together with technical details were described in Refs.~\cite{Hou:2016sho,Pumplin:2001ct,Martin:2002aw,Martin:2009iq,Khanpour:2016pph,Shoeibi:2017lrl}.
The most common and effective approach is the ``Hessian method''. In this section the outline of this method is explained because it is used in our analysis.
Up to the leading quadratic terms, the increase $\chi^2$ can be written in terms of the Hessian matrix
\begin{equation}\label{eq:hij}
H_{ij} \equiv \frac{1}{2} \frac{\partial^2 \chi^2}{\partial a_i \partial a_j} \Bigg|_{\rm min}
\end{equation}
as
\begin{equation}\label{eq:chi2hessian}
\Delta \chi_{\mathrm {global}}^2 = \chi^2_{\rm global} - \chi^2_{\rm min} = \sum_{ij}^{N_{\mathrm{par}}} H_{ij} (a_i-a_i^{(0)}) (a_j-a_j^{(0)}),
\end{equation}
where $a_i$ ($i=$1, 2, ..., N) refers to the free parameters for each PDF presented in Eq.~(\ref{input_PDFs}) and N is the number of parameters. $\Delta \chi_{\rm global}^2$ illustrates the allowed variation in $\chi^2$. The standard formula for linear error propagation is
\begin{equation}\label{eq:1uncertainties1}
(\Delta q)^2 = \Delta \chi^2_{\mathrm {global}}  \sum_{i , j}  \frac{\partial \Delta q}{\partial a_i} (H_{i j})^{-1} \frac{\partial \Delta q}{\partial a_j}  \,   .
\end{equation}
The parameter value of polarized PDF, i.e. ${a_1^0,...,a_n^0}$, extracted from the NNLO QCD fit, will be presented in Sec.~\ref{results}.
Equation~(\ref{eq:1uncertainties1}) is not convenient to use since the derivative of $\Delta q$ with respect to each parameter $a_i$ is required. An improved iterative method has been devised in \cite{Pumplin:2001ct,Martin:2002aw,Martin:2009iq} in which the Hessian matrix is diagonalized.
We adopt this improved Hessian method in our analysis and work in terms of rescaled eigenvectors and eigenvalues.
The PDF uncertainty determination becomes much easier in terms of the appropriately normalized eigenvectors, $z_i$
\begin{equation}\label{eq:deltachi2z}
\Delta \chi_{\rm {global}}^2 = \sum_i^{N_{\mathrm{par}}} z_i^2\,.
\end{equation}
The uncertainty of an individual PDF at particular values of $x$ and $Q^2$ can be estimated using
\begin{equation}\label{eq:uncertainties-1}
\Delta q = \Delta \chi^2_{\rm global}\frac{1}{2}
\sqrt{\sum_{k=1}^n [ \Delta q(S_k^+)-\Delta q(S_k^-)]^2} \,.
\end{equation}
Here, $S_k^+$ and $S_k^-$ are polarized PDF sets displaced along eigenvector directions by the extracted $\Delta \chi_{\rm global}^2$ value.

With the standard ``parameter-fitting'' criterion, we would expect the uncertainties to be given by the choice of tolerance parameter
$T^2 = \Delta \chi_{\rm {global}}^2 = 1$ for the 68\% (one-sigma-error) confidence level (C.L.) limit~~\cite{Martin:2009iq,Martin:2002aw}.
For the general case with N degrees of freedom, the $\Delta \chi^2_{\rm {global}}$ value needs to be calculated to determine the
size of the uncertainties.
Assuming that $\Delta \chi^2_{\rm global}$ follows the $\chi^2$ distribution with N degrees of freedom, we have the C.L. $P$ as~\cite{Martin:2009iq,Martin:2002aw,Hirai:2003pm}
\begin{equation}\label{eq:uncertainties-2}
P = \int_0^{\Delta \chi^2_{\rm {global}}}\frac{1}{2\Gamma(\rm N/2)(\frac{S}{2})^{(\rm N/2)-1}} e^{-\frac{S}{2}} \, dS \,.
\end{equation}
%
In the case of the one-free-parameter fit, one obviously has $\Delta \chi^2_{\rm {global}} = 1$.
Since the polarized parton distributions in common QCD fits are considered with several free parameters, $N > 1$, the value of $\Delta \chi^2_{\rm {global}}$ should be reevaluated form Eq.~(\ref{eq:uncertainties-2}).
Here we calculate the uncertainties of polarized PDFs with $\Delta \chi^2_{\rm {global}} = 1$. For other values of $\Delta \chi^2_{\rm {global}}$, one can simply scale our error bands by $(\Delta \chi^2_{\rm {global}})^{1/2}$.

\section{Polarized structure functions for nuclei }\label{Nuclear-polarized-structure-functions}

The experimental data on deep inelastic lepton-nucleus ($\ell^\pm + A$) scattering can reveal more information on the behavior of spin structure functions of nucleon and nucleon correlations in nuclei at small and large values of Bjorken $x$. It will provides more reliable pictures of the nuclear phenomena. Moreover, the absence of free neutron targets means that polarized light nuclei such as deuterium and $^3{\mathrm He}$ must be used as effective polarized neutron targets.
The scattering of polarized lepton beams $\ell^\pm$ from polarized nuclear targets paves the way to detail study of spin structure of the nucleon encoded in the spin structure functions of $g_1 (x, Q^2)$ and $g_2 (x,Q^2)$.
In order to extract these spin structure functions from the spin-dependent DIS data on nuclear targets, one needs to account for the nontrivial nuclear corrections.
The nuclear effects that play an important role in the polarized as well as unpolarized DIS on nuclei can be divided into ``coherent'' and
``incoherent'' contributions~\cite{Bissey:2001cw,Guzey:2000wh}. Incoherent nuclear effects, which are present at all values of Bjorken $x$, arise from the scattering of the incoming leptons on each individual nucleons.
Some examples of incoherent nuclear effects are the well-known Fermi motions, spin depolarizations, binding and the presence of a non-nucleonic degree of freedom.

The coherent nuclear effects are typically important at small values of the momentum fraction $x$. They result from the interaction of the incoming leptons with two or more nucleons inside the targets.
Nuclear shadowing which is very important for the region $10^{-4} \leq x \leq 0.03-0.07$, and nuclear anti-shadowing at $0.03-0.07 \leq x \leq 0.2$ are examples of the coherent nuclear effects. Nuclear shadowing corrections in polarized $^3{\mathrm He}$ have been argued in detail in Refs.~\cite{Frankfurt:1996nf,Guzey:1999rq}. These effects arise from multiple scattering of the leptons from two or more nucleons in the $^3{\mathrm He}$ nucleus. In addition to the point mentioned above, contributions to the polarized $^3{\mathrm He}$ structure function from non-nucleonic degrees of freedom in the nucleus have been discussed in detail in Refs.~\cite{Bissey:2001cw,Guzey:2000wh}.

For the case of the polarized DIS in which we present in the following analysis, the major contributions come from the incoherent scattering on the nucleons of the targets.
The aim of this work is to study these nuclear effects for the case of the polarized DIS at NNLO approximation and assess their importance on the spin structure functions of  $g_{1,2}^{^3{\mathrm He}}$ and $g_1^{^3{\mathrm H}}$.

In following, we describe the nuclear effects in inclusive scattering of polarized leptons from polarized helium $^3{\mathrm He}$ and tritium $^3{\mathrm H}$ nuclei focusing in particular on kinematics at whole values of momentum fraction $x$.

%
\subsection{  Weak binding approximation  (WBA) }
%

To begin our discussion on the nuclear corrections, we should note that in the standard nuclear structure function analyses, the contributions from the spin depolarization, Fermi motions and binding are described within the framework of the so-called ``convolution approach''.
The polarized nuclear structure function of $^3{\rm He}$ and $^3{\mathrm H}$ can be written as a convolution of the nucleons light-cone momentum distributions with the off-shell polarized nucleon structure functions of the protons $g_1^p (x, Q^2)$ and neutrons $g_1^n (x, Q^2)$.
In what follows, we apply the method presented in Refs.~\cite{Bissey:2001cw,Ethier:2013hna} and adopt the approach in which the nucleons light-cone momentum distributions can be extracted from the nuclear spectral functions~\cite{Bissey:2000ed,Afnan:2003vh}.

The well-known coherent effects associated with the multiple scattering from two or more nucleons inside the nucleus give rise to corrections at small values of Bjorken $x$.
In order to study the nuclear effects in the incoherent scattering, we restrict ourselves to the intermediate and large regions of momentum fraction $x$ in which the incoherent scattering from a single nucleon is assumed to dominate~\cite{Piller:1999wx,FernandezdeCordoba:1995pt,Ethier:2013hna,Jimenez-Delgado:2013boa}.
In this framework, in which the nucleus is treated as a non-relativistic system of weakly bound nucleons, the spin-dependent structure functions of $^3{\mathrm He}$ can be
obtained as~\cite{Kulagin:1994fz,Kulagin:1994cj,Kulagin:2007ph,Kulagin:2008fm,Kulagin:2004ie,CiofidegliAtti:1993zs}
\begin{eqnarray}\label{g13He-1}
g_{1}^{^3{\rm He}}(x, Q^{2})  =
\int \frac{dy}{y} \Big[ 2 \Delta f^{p} 
(y, \gamma) g_{1}^{p}(\frac{x}{y}, Q^{2})  
& + &  \nonumber   \\
\Delta f^{n} (y, \gamma) 
g_{1}^{n} (\frac{x}{y}, Q^{2}) \Big] \,,
\end{eqnarray}
where $y = \frac{p.q}{{\rm M} \nu}$ is the nuclear light-cone momentum fraction carried by the interacting
nucleons inside the nucleus. The functions $f^{N(=p, n)} (y, \gamma)$ are the nucleon light-cone momentum distributions in the $^3{\mathrm He}$ nucleus computed in terms
of the nuclear spectral functions~\cite{Bissey:2001cw,Ethier:2013hna}
\begin{eqnarray}\label{spectral-function}
\Delta f^{\rm N} (y, \gamma)  =  \int \frac{d^{4} p}
{(2 \pi)^{4}} D^{N}(\varepsilon, p, \gamma)
\, \delta(y - 1 -\frac{\varepsilon \,  
+
\,  \gamma p_{z}}
{\rm M}) \,, \nonumber   \\
\end{eqnarray}
with N = proton or neutron, and $D^{N}$ is the energy-momentum distribution functions.
In the Bjorken limit in which $(\gamma \to 1)$, the nucleon light-cone momentum distributions $f^{\rm N} (y, \gamma)$ depend only to the 
$y$ variables, which covers the range between $x$ and $\frac{{\rm M}_{^3{\mathrm He}}}{\rm M} \approx 3$.
Finally, the polarized structure functions of $g_{1}^{^3{\mathrm He}} (x, Q^2)$ and $g_{1}^{^3{\rm H}} (x, Q^2)$ can be written as
\begin{eqnarray}\label{g13He}
g_1^{^3{\rm He}}(x,Q^2)  = \int_x^3 \frac{dy}{y}
 \Delta f^n_{^3{\mathrm He}}(y) 
g_1^n(x/y, Q^2) & + &  \nonumber   \\
2 \int_x^3 \frac{dy}{y}
\Delta f^p_{^3{\mathrm He}}(y)g_1^p(x/y, Q^2) \,,
\end{eqnarray}
\begin{eqnarray}\label{g13H}
g_{1}^{^3{\rm H}}(x, Q^{2}) = 2 \int_{x}^{3}
 \frac{dy}{y} \Delta f^{n}_{^3{\mathrm H}}(y) 
g_1^n(x/y, Q^{2}) & + &  \nonumber   \\
\int_{x}^{3} \frac{dy}{y}
\Delta f^p_{^3{\mathrm H}}(y)g_1^p(x/y, Q^{2}) \ .
\end{eqnarray}
The Fermi motions and their bindings are parametrized
through the distributions of $\Delta f^{N (=p, n)}_{^3{\mathrm He}}$ and $\Delta f^{N (=p, n)}_{^3{\mathrm H}}$ which can be calculated using the ground-state
wave functions of $^3{\mathrm He}$ and $^3{\mathrm H}$ nuclei, respectively.
It is worth mentioning here that, because of isospin symmetry, the light-cone momentum distribution $\Delta f^p_{^3{\mathrm He}}(y)$ is equal to the $\Delta f^n_{^3{\mathrm H}}(y)$.

\begin{figure}[H]
\includegraphics[clip,width=0.50\textwidth]{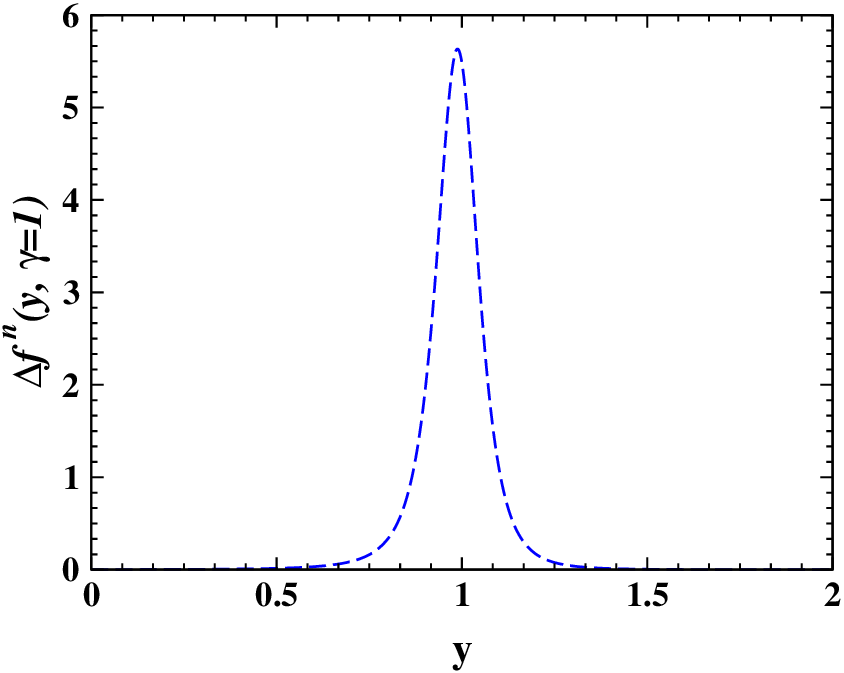}
\begin{center}
\caption{{\small The polarized neutron light-cone momentum distribution functions at the Bjorken limit  $(\gamma \to 1)$ in $^3{\mathrm He}$, based on the results presented in  Ref.~\cite{Bissey:2000ed}. \label{fig:specteralnoutron}}}
\end{center}
\end{figure}

\begin{figure}[H]
\includegraphics[clip,width=0.50\textwidth]{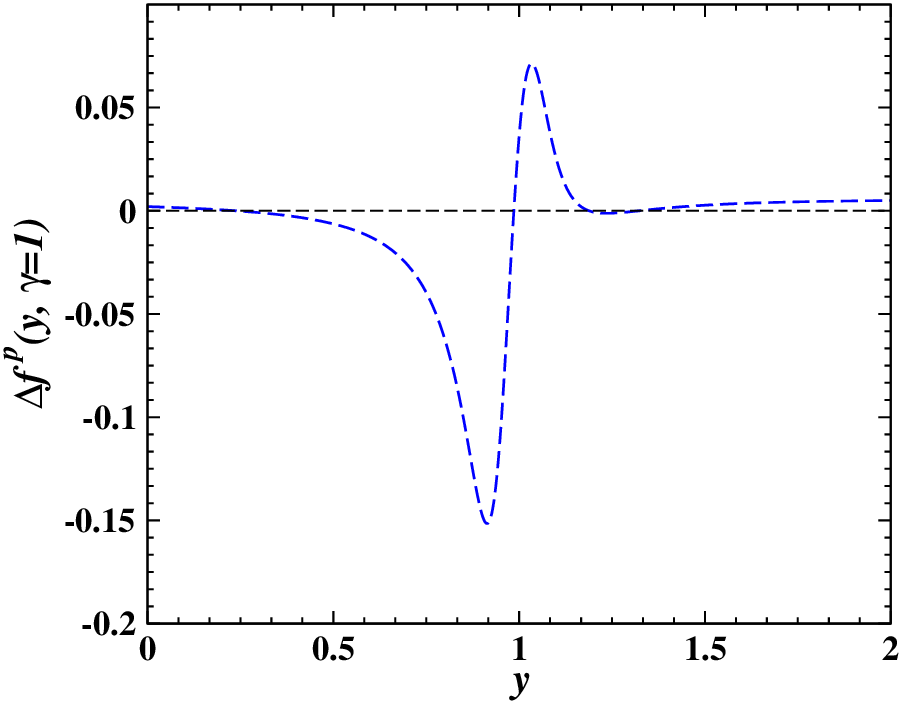}
\begin{center}
\caption{{\small The polarized proton light-cone momentum distribution functions at the Bjorken limit $(\gamma \to 1)$ in $^3{\mathrm He}$, based on the results presented in 
Ref.~\cite{Bissey:2000ed}. \label{fig:specteralproton}}}
\end{center}
\end{figure}

In order to obtain the polarized light-cone momentum distributions, we used the numerical results presented in Ref.~\cite{Bissey:2000ed}.
The polarized light-cone distribution functions for the neutron $\Delta f^{n}_{^{3}{\mathrm He}}(y)$ as well as for the proton $\Delta f^{p}_{^{3}{\mathrm He}}(y)$ are shown as a function of $y$ in Figs.~\ref{fig:specteralnoutron} and \ref{fig:specteralproton}, respectively. It has been shown in various studies that the polarized light-cone distribution functions $\Delta f^N(y, \gamma=1)$ are sharply peaked at $y \approx 1$~\cite{Bissey:2000ed,Afnan:2003vh,Kulagin:2008fm}.
%
%

\subsection{Effective polarizations of the nucleons}

In the limit of zero nuclear binding, and in the Bjorken limits in which $(\gamma \to 1)$, the polarized proton light-cone momentum distribution functions $\Delta f^{\rm N}(y)$ become infinitesimally narrower and are sharply peaked around the points $y \approx 1$ \{$f^{\rm N} \sim \delta (1 - y)$\}, due to the small average separation energy per nucleons inside the nuclei.

Thus, in this approximation, one can express the polarized nuclear structure functions $g_{1}^{A} (x, Q^2)$ as linear combinations of the
polarized proton $g_1^{p} (x, Q^{2})$ and neutron $g_1^n (x, Q^2)$ structure functions weighted by effective polarizations.
Consequently, Eq.~(\ref{g13He-1}) is often approximated by
\begin{eqnarray}
g_{1}^{^3{\mathrm He}}
(x, Q^{2}) =
2 P^p g_{1}^p(x, Q^{2})  
+
P^n g_{1}^{n} (x, Q^{2})\ .
\end{eqnarray}
Here $P^{p}$ and $P^{n}$ are the proton and neutron effective polarizations inside the polarized $^3{\mathrm He}$ nucleus.
The proton effective polarizations $P^{p}$ are assumed to be the average polarizations of the two protons inside the $^3{\mathrm He}$ nuclei.
The effective polarizations can be described in terms of integrals of the diagonal light-cone momentum distribution functions at the Bjorken limit in which $(\gamma \to 1)$,
\begin{eqnarray}
P^{\rm p} = \int_0^{3} dy \, 
\Delta f^p_{^3{\mathrm He}} 
(y, \gamma=1)  \,,  \nonumber   \\
P^{\rm n} = \int_0^{3} dy \,
\Delta f^n_{^3{\mathrm He}}
(y, \gamma=1)  \,.
\end{eqnarray}
The effective polarizations presented in our analysis can be computed numerically from models of the $^3{\mathrm He}$ nucleus wave functions.
The calculations of Refs.~\cite{Bissey:2000ed,Friar:1990vx} have shown that $P^{p} = -0.028 \pm 0.004$ and $P^{n} = 0.86 \pm 0.02$.

\subsection{Non-nucleonic contributions}

As we have discussed earlier, free nucleons behave differently from nucleons in which bound in the nuclei.
This is due to the some important nuclear corrections such as the nuclear bindings, Fermi motions, nuclear shadowing, and anti-shadowing as well as the non-nucleonic degrees of freedom.
Consequently, the descriptions of the nucleon as a mere collections of the protons and neutrons (descriptions of nuclear properties in terms of the nucleon degrees
of freedom alone) may not be complete, and, hence the nuclear corrections due to the non-nucleonic degrees of freedom have to be considered.
For the spin-dependent observables $g_{1}^{A}(x, Q^{2})$, small admixtures of the $\Delta (1232)$ isobar in three-body wave functions were
found to be necessary for a better descriptions of the polarized nuclear structure functions~\cite{Saito:1990aj}.

In order to consider the nuclear effects in which originate from the non-nucleonic degrees of freedom, we utilized the works presented in Ref.~\cite{Bissey:2001cw} which provides a description of the $g_{1} (x, Q^{2})$ spin structure functions of helium $^3{\mathrm He}$ as well as tritium $^3{\mathrm H}$ nuclei over the range of $10^{-4} \leq x  \leq 0.8$.
It models the $^3{\mathrm He}$ wave function including the $S$, $S^\prime$, and $D$ states, and the non-nucleonic degrees of freedoms from the effect of $\Delta (1232)$ isobar.
The corresponding results for the $^3{\mathrm He}$ and $^3{\mathrm H}$ can be written,
\begin{eqnarray}\label{g_1He-degrees-of-freedom}
&& g_1^{^3{\mathrm He}} = \int_x^{3} \frac{dy}{y}
\Delta f^{n}_{^3{\mathrm He}}(y) g_1^{n}(x/y)
+ 2 \int_x^{3} \frac{dy}{y}
\Delta f^{p}_{^3{\mathrm He}}(y) g_{1}^{p}(x/y) \nonumber  \\
&& - 0.014 \Big(g_{1}^{p}(x)
- 4 g_{1}^{n}(x) \Big) \,,
\end{eqnarray}
and
\begin{eqnarray}\label{g_1H-degrees-of-freedom}
&&g_1^{^3{\rm H}} = 
2 \int_x^3 \frac{dy}{y}
\Delta f^n_{^3{\mathrm H}}(y)  g_{1}^{p} (x/y)
+ \int_x^3 \frac{dy}{y}
\Delta f^p_{^3{\mathrm H}}(y)
g_1^n (x/y)    \nonumber \\
&& + 0.014 \Big(g_{1}^{p} (x)
- 4 g_{1}^{n} (x)  \Big)  \,.
\end{eqnarray}
The last terms in Eqs.~\eqref{g_1He-degrees-of-freedom} and \eqref{g_1H-degrees-of-freedom} arise
from the $\Delta (1232)$ component in the $^3{\mathrm He}$ and $^3{\mathrm H}$ wave functions.
They will have sizable contributions at large values of Bjorken $x$, $0.2 \leq x \leq 0.8$.

The same formula can be applied for the $g_2^A$ nuclear structure functions as
\begin{eqnarray}
&& g_2^{^3{\rm He}} = \int_x^{3} \frac{dy}{y} \Delta 
f^n_{^3{\rm He}}(y) g_2^{n}(x/y)
+ 2 \int_x^3 \frac{dy}{y} \Delta
 f^p_{^3{\rm He}}(y) g_2^{p} (x/y)  \nonumber \\
&& -0.014 \Big( g_{2}^p(x) -
 4 g_2^n (x) \Big) \,,
\end{eqnarray}
and
\begin{eqnarray}
&&g_2^{^3{\rm H}} = 2 \int_x^3
 \frac{dy}{y} \Delta f^{n}_{^3{\rm H}}(y) g_2^p (x/y)
+ \int_x^3 \frac{dy}{y}
 \Delta f^p_{^3{\rm H}}(y) g_2^{n} (x/y)  \nonumber \\
&& + 0.014 \Big( g_{2}^p(x) -
 4 g_2^n (x)\Big)  \,.
\end{eqnarray}

\subsection{Shadowing and anti-shadowing corrections}

At high energy or small values of momentum fraction $x$, the virtual photons can interact ``coherently'' with several nucleons in the nuclear targets.
These behaviors are manifested in the nuclear shadowing and anti-shadowing effects, and breaks down the well-known convolution approximations~\cite{Bissey:2001cw}.
Considering these nuclear corrections, one can write the polarized nuclear structure functions as
\begin{eqnarray}\label{g1Hesh}
&& g_1^{^3{\mathrm He}} = \int_x^3 \frac{dy}{y}
 \Delta f^{n}_{^3{\mathrm He}}(y) g_1^{n}(x/y)
+2 \int_x^3 \frac{dy}{y} \Delta 
f^p_{^3{\mathrm He}}(y) g_1^{p}(x/y)  \nonumber \\
&& - 0.014 \Big( g_1^{p}(x) -
 4 g_1^{n}(x)\Big) + a(x) g_1^{n}(x) 
 + b(x) g_1^p(x)  \ ,\nonumber\\
\end{eqnarray}
and
\begin{eqnarray}\label{g1Hsh}
&&g_1^{^3{\mathrm H}} = 2 \int_x^{3} 
\frac{dy}{y} \Delta f^n_{^3{\mathrm H}}(y) g_1^{p}(x/y)
+ \int_x^3 \frac{dy}{y} \Delta
 f^p_{^3{\mathrm H}}(y) g_1^{n} (x/y)  \nonumber \\
&& + 0.014 \Big( g_1^{p}(x) - 
4 g_1^{n} (x) \Big) + a(x) g_1^{n}(x)
 + b(x) g_1^{p}(x)  \ ,\nonumber\\
\end{eqnarray}
The functions $a(x)$ and $b(x)$ describe both the corrections of nuclear shadowing and anti-shadowing, and are functions of $x$ and $Q^2$.

Since in the most nuclear polarized DIS experiments the $x$ coverage does not drop below the limits $x \sim 0.2$, the shadowing and anti-shadowing corrections can be ignored~\cite{Flay:2016wie}.
However the calculations of Ref.~\cite{Bissey:2001cw} have shown that these two effects are quite significant and can affect the extraction of the nucleon spin functions at small values of Bjorken $x$.
As we mentioned earlier, current experimental data do not reach to very small values of $x$.
Consequently, the corrections from shadowing ($10^{-4} \leq x \leq 0.03-0.07$) and anti-shadowing ($0.07-0.03 \leq x \leq 0.2$) can be completely ignored in the analysis of the present DIS data on polarized nuclei.

\section{Discussion of fit results}\label{results}

The best values for parton distribution functions are demonstrated in Table~\ref{TablePPDfs}. Parameters marked with ($^*$) are fixed after
an initial minimization step to their best values.
Accordingly, there are 16 unknown parameters,including the strong coupling constant, that provide enough flexibility to have a reliable fit. We achieve $\chi^2/dof=473.195/495=0.955$,that provides an acceptable fit to data.
We extract the strong coupling constant simultaneously with polarized PDF parameters to study its correlation with the others. We obtain the value of $\alpha_s(Q_0^2)=0.30998\pm 0.0113$ at a $0.68\%$ confidence level. Rescaling the coupling constant to the mass of the Z boson, we achieve $\alpha_s(M_Z^2)=0.1106\pm 0.0010$. The present world average value is $\alpha_s(M_Z^2)=0.1185 \pm 0.0006$~\cite{Olive:2016xmw}.
%
\begin{table*}[htb]
	\begin{center}
		\caption{  \label{TablePPDfs} Obtained parameter values and their statistical
			errors at the input scale Q$_0^2$ = 1 GeV$^2$ determined from  leading-twist analyses in the NNLO approximation. Those marked with ($^*$) are fixed. Note that the TMCs are included in our QCD analysis.}
		\begin{tabular}{c c c c c c}
			\hline  \hline
			{Flavor} &  $\eta$ &  $\alpha$  &  $\beta$ &  $\epsilon$ &  $\gamma$
			\\
			\hline\hline
			$u+\bar{u}$ & $~0.807^*~ $ &$~ 0.259\pm 0.007~$& $~2.857 \pm 0.049~$ & $~-4.95\pm 0.324~$ & $~38.12\pm1.85~$ \\
			$d+\bar{d}$ & $~-0.461^*~$  &$~0.332 \pm 0.72~ $ & $~3.139 \pm 0.75~$ &$~0~$& $~4.53\pm 1.85~$  \\
			$s+\bar{s}$ & $~-0.119 \pm 0.008~$ & $~0.249 \pm 0.048~$ & $~15.68\pm  4.21~$  & $~0~$ & $~0~$   \\
			$G$			& $~0.133 \pm 0.027~$  &$~23.33 \pm 4.32~$&$~86.57 \pm 6.35~$ & $~1.434 \pm 0.23~$ & $~-4.992 \pm 0.574~$ \\  \hline
			\hline
		\end{tabular}
	\end{center}
\end{table*}

%
%
\subsection{Polarized parton distribution functions}

\begin{figure}[!htb]
	\vspace*{0.50cm}
	\includegraphics[clip,width=0.50\textwidth]{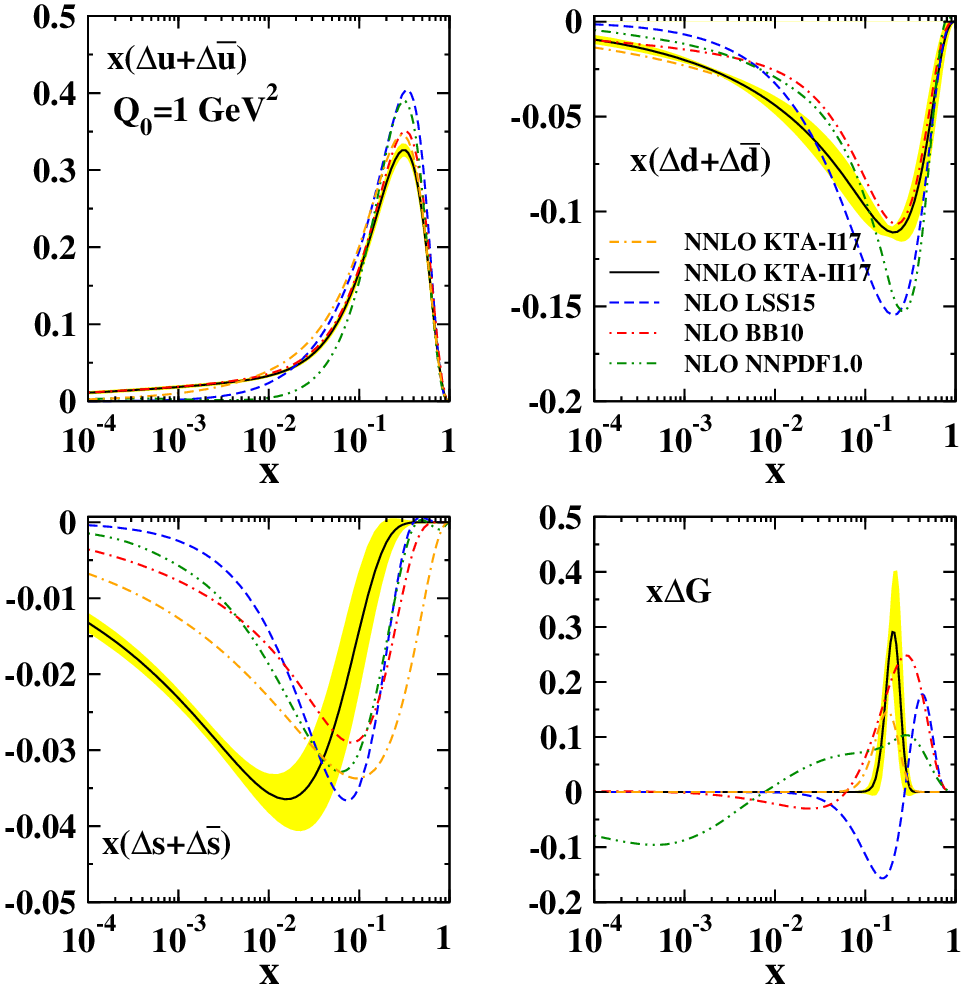}
	\begin{center}
		\caption{{\small  Our results KTA-II17 for the polarized PDFs at Q$_0^2=$ 1 GeV$^2$ as a function of $x$ in the NNLO approximation plotted as a solid curve along with their $68\%$ C.L. uncertainties, as described in the text.
				We also show the results obtained in earlier global analyses of
				KTA-I17 (dashed-dashed-dotted)~\cite{Khanpour:2017cha},
				LSS15 (dashed)~\cite{Leader:2014uua},
				BB10 (dashed-dotted)~\cite{Blumlein:2010rn},
				NNPDF1.0 (dashed-dotted-dotted)~\cite{Nocera:2014gqa}. \label{fig:partonQ0polarizeTMC}}}
	\end{center}
\end{figure}

Our NNLO polarized PDF along with their corresponding $0.68\%$ C.L. uncertainties are presented in Fig.~\ref{fig:partonQ0polarizeTMC}. Various parametrizations from the literature~ \cite{Leader:2014uua,Blumlein:2010rn,Nocera:2014gqa} and \cite{Khanpour:2017cha} obtained from NLO and NNLO QCD analyses of the inclusive data are presented for comparison.
The $x(\Delta u+\Delta \bar{u})$ and $x\Delta G$ distributions are positive, while the $x(\Delta d + \Delta \bar{d})$ and $x(\Delta s + \Delta \bar{s})$ distributions are negative.

For the $x(\Delta u+\Delta \bar{u})$ distributions, all of the curves are comparable. Examining the $x(\Delta d+\Delta \bar{d})$ distributions we see that most of the fits are in agreement, with the possible exception of the LSS15 and NNPDF1.0.

All analyses of the polarized inclusive DIS data have extracted significantly negative results for the polarized strange quark distribution functions, $x(\Delta s+\Delta \bar{s})$, for all $x$ values, even though the parametrization allowed a sign change as a function of $x$.

Results for polarized gluon distribution from the various fits on the present polarized inclusive DIS data are quite spread. The difficulties in constraining $x\Delta G$ cannot be ruled out with present data. Our gluon distribution tends to zero more quickly than the others.

%
%
\subsection{$g_1$ structure functions}

\begin{figure*}[!htb]
	\includegraphics[clip,width=0.45\textwidth]{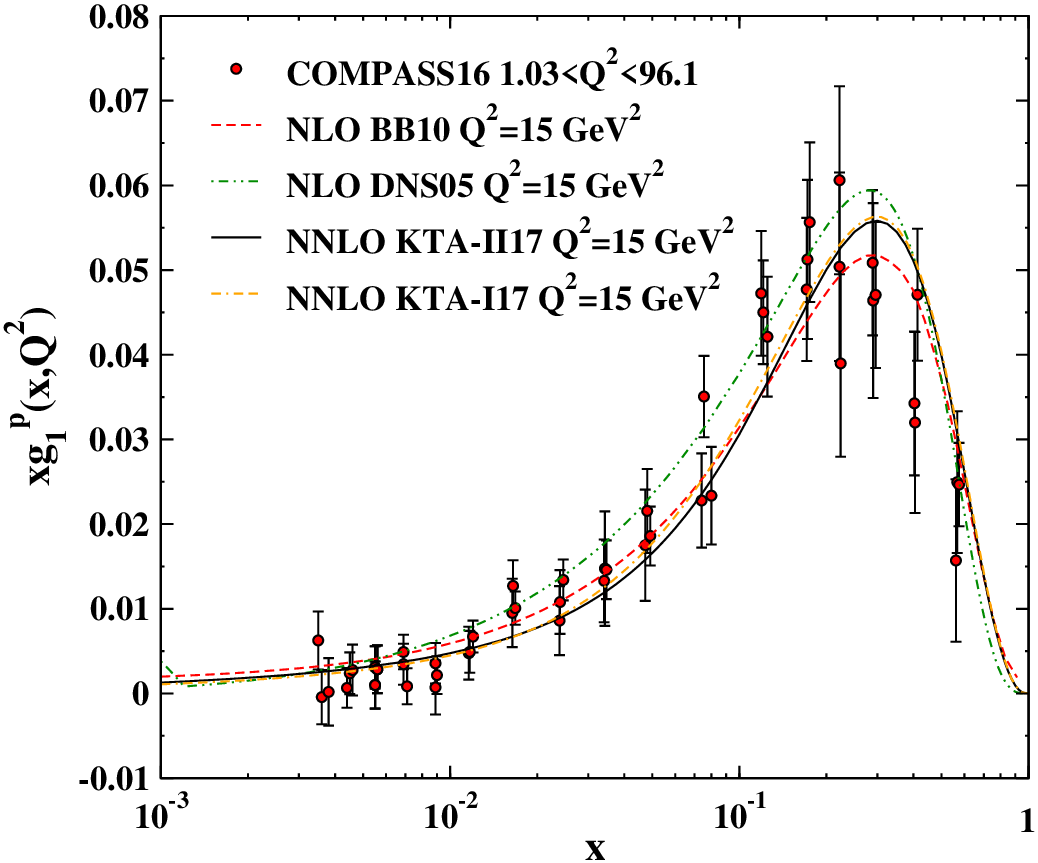}
	\includegraphics[clip,width=0.45\textwidth]{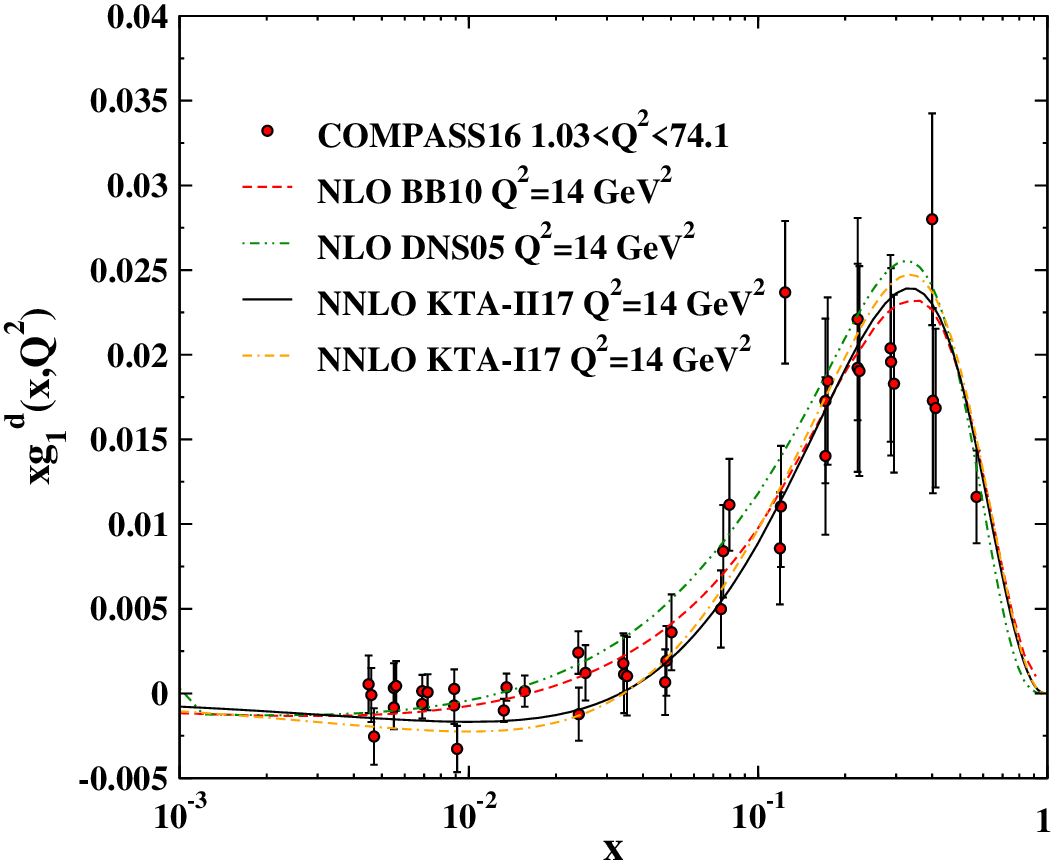}
	\begin{center}
		\caption{{\small  KTA-II17 prediction (solid curve) for the polarized structure function of the proton (left) and deuteron (right) as a function of $x$ for the mean value of $Q^2=15$ GeV$^2$ and $Q^2=14$ GeV$^2$, respectively. Also shown are BB10 (dashed) \cite{Blumlein:2010rn}, DNS05 (dashed-dotted-dotted) \cite{deFlorian:2005mw} extracted at the NLO approximation, and KTA-I17 (dashed-dashed-dotted)~\cite{Khanpour:2017cha} obtained at the NNLO approximation together with the recent experimental data from the COMPASS16 collaborations ~\cite{Adolph:2015saz,Adolph:2016myg}.  \label{fig:xg1p}}}
	\end{center}
\end{figure*}

Figure~\ref{fig:xg1p} represents results for the polarized structure function $xg_1^p$ and $xg_1^d$  as a function of $x$ at $Q^2=$ 15 GeV$^2$ and $Q^2=$ 14 GeV$^2$, respectively.
For comparison, we illustrate the results extracted in
BB10~\cite{Blumlein:2010rn}, DNS05~\cite{deFlorian:2005mw} at the NLO approximation, and KTA-I17~\cite{Khanpour:2017cha} at the NNLO approximation.
Our curves stay compatible with the recent COMPASS16 published data~\cite{Adolph:2015saz,Adolph:2016myg} within statistical uncertainties.
Note that the experimental observables belong to the scale region of $1.03<Q^2<96.1$ GeV$^2$ and $1.03<Q^2<74.1$ GeV$^2$ for proton and deuteron polarized structure functions.

\begin{figure}[!htb]
	\includegraphics[clip,width=0.45\textwidth]{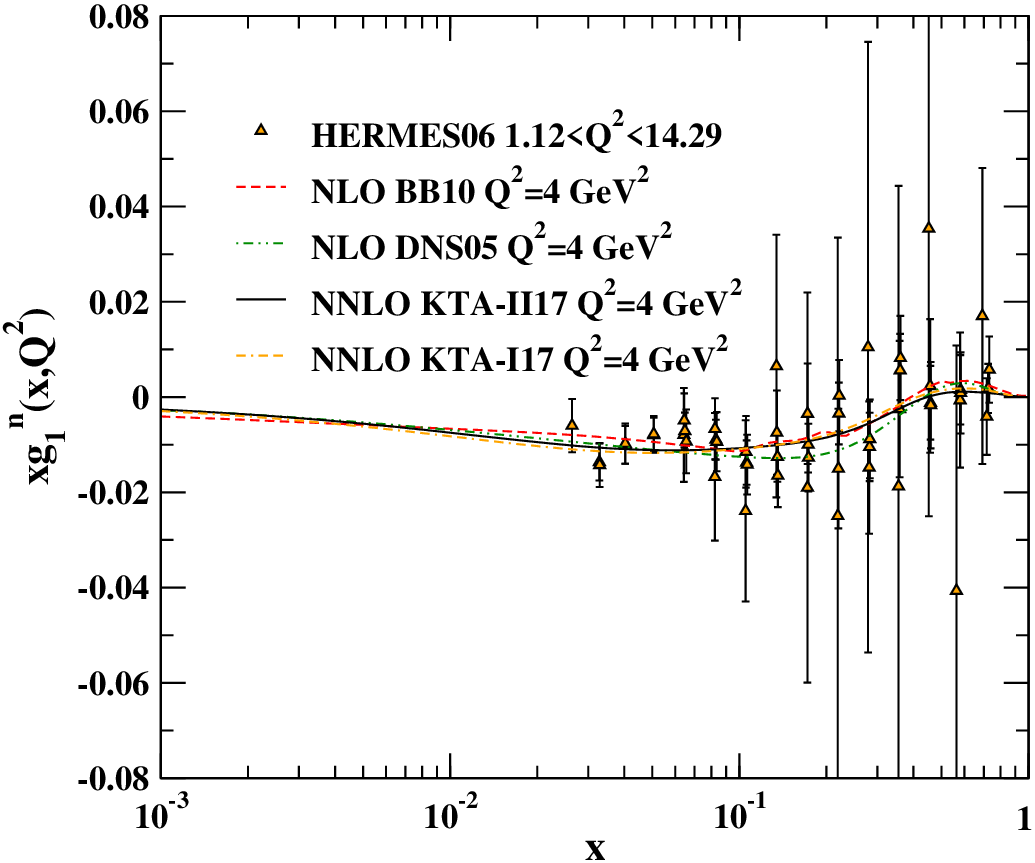}
	\begin{center}
		\caption{{\small  KTA-II17 prediction (solid curve) for the polarized structure function of neutron for the mean value of $Q^2=4$ GeV$^2$ as a function of $x$ in the NNLO approximation. Also shown are BB10 (dashed)~\cite{Blumlein:2010rn},
				DNS05 (dashed-dotted-dotted)~\cite{deFlorian:2005mw} extracted at the NLO approximation, and KTA-I17 (dashed-dashed-dotted)~\cite{Khanpour:2017cha} obtained at the NNLO approximation together with the experimental data from the HERMES06 Collaboration~\cite{Ackerstaff:1997ws}.  \label{fig:xg1n}}}
	\end{center}
\end{figure}

In Fig.~\ref{fig:xg1n} we present our result for the polarized structure function of neutron $xg_1^n$ as a function of $x$ at $Q_0^2=$ 4 GeV$^2$.
We observe that our result coincides with those of BB10~\cite{Blumlein:2010rn}, DNS05~\cite{deFlorian:2005mw} at the NLO approximation, and KTA-I17~\cite{Khanpour:2017cha} at the NNLO approximation.
The HERMES06 experimental data~\cite{Ackerstaff:1997ws} are well described within errors by all the curves.  The experimental observables belong to the scale region of $1.12<Q^2<14.29$ GeV$^2$.
Generally the $xg_1^n$ data have larger uncertainties compared with the $xg_1^p$ and $xg_1^d$ data.
More accurate experimental measurements on light nuclear targets are required to allow us to scrutinize neutron structure functions.

%
%
\subsection{Nuclear polarized structure functions }\label{Numerical_results}

We are in a position to apply the formalism developed in Sec.~\ref{Nuclear-polarized-structure-functions} to
compute the $g_{1,2}^{^3{\mathrm He}}$ and $g_{1}^{^3{\mathrm H}}$ structure functions and corresponding nuclear corrections based on the extracted PPDFs.
In particular, we study the impact of nuclear effects originating from the ``non-nucleonic degrees of freedom'' on the extraction of the spin structure of
the $^3{\mathrm He}$ and $^3{\mathrm H}$.

\begin{figure}[htb]
	\includegraphics[clip,width=0.450\textwidth]{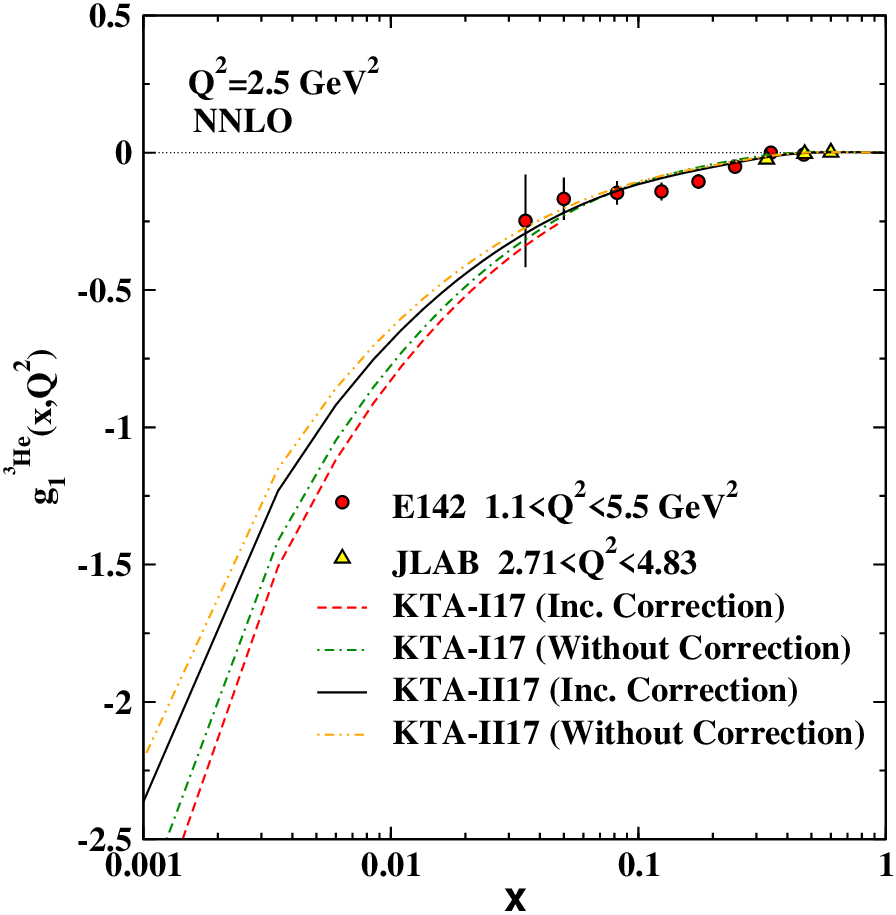}
	\begin{center}
		\caption{{\small  Analytical results  with and without nuclear corrections for the polarized structure function of $g_1^{^3{\mathrm He}} (x, Q^2)$ as a function of $x$ at NNLO approximation. The current fit is the solid curve. Also shown are  data form E142~\cite{Anthony:1996mw} and JLAB04~\cite{Zheng:2004ce}. \label{fig:g1He3}}}
	\end{center}
\end{figure}

\begin{figure}[htb]
	\includegraphics[clip,width=0.450\textwidth]{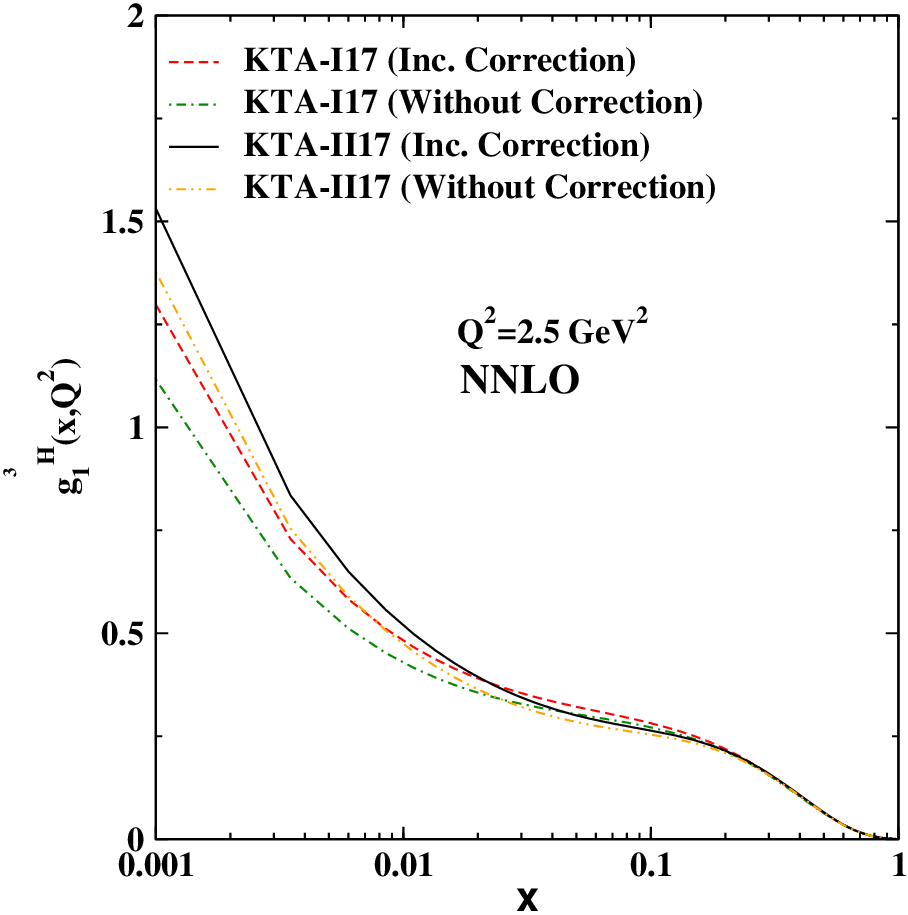}
	\begin{center}
		\caption{{\small  Analytical results  with and without nuclear corrections for the polarized structure function of $g_1^{^3{\mathrm H}} (x, Q^2)$ as a function of $x$ at NNLO approximation. The current fit is the solid curve. Also shown are the QCD NNLO curves obtained by KTA-I17~\cite{Khanpour:2017cha} for comparison. \label{fig:g1H3}}}
	\end{center}
\end{figure}

In Figs.~(\ref{fig:g1He3}) and~(\ref{fig:g1H3}), we show our results for the $g_1^{^3{\mathrm He}}$ and $g_1^{^3{\mathrm H}}$ polarized structure functions at NNLO approximation based on Eqs.~(\ref{g_1He-degrees-of-freedom}) and ~(\ref{g_1H-degrees-of-freedom}) respectively, and compare with the curves based on KTA-I17~\cite{Khanpour:2017cha}.
The experimental data from E142~\cite{Anthony:1996mw} and JLAB04~\cite{Zheng:2004ce} are well described by the fit in Fig.~\ref{fig:g1He3}.
One can conclude that our results for the $g_1^{^3{\mathrm He}}$ based on both of our spin-dependent PDFs (KTA-I17 and KTA-II17) reproduce the trend of the existing data.

We would like to stress again that the small $x$ ($10^{-4} \leq x \leq 0.2$) effects from nuclear shadowing and antishadowing were not taken into account in the results presented in Figs.~\ref{fig:g1He3} and \ref{fig:g1H3}. At large $x$ values, $x \geq 0.8$, all the results coincide. The nuclear corrections lead to a sizable difference in the small $x$ values.
For intermediate and low $x$ values, including nuclear corrections underestimates the results.

%
%

\begin{figure*}[htb]
	\begin{center}
		\vspace{0.50cm}
		\resizebox{0.45\textwidth}{!}{\includegraphics{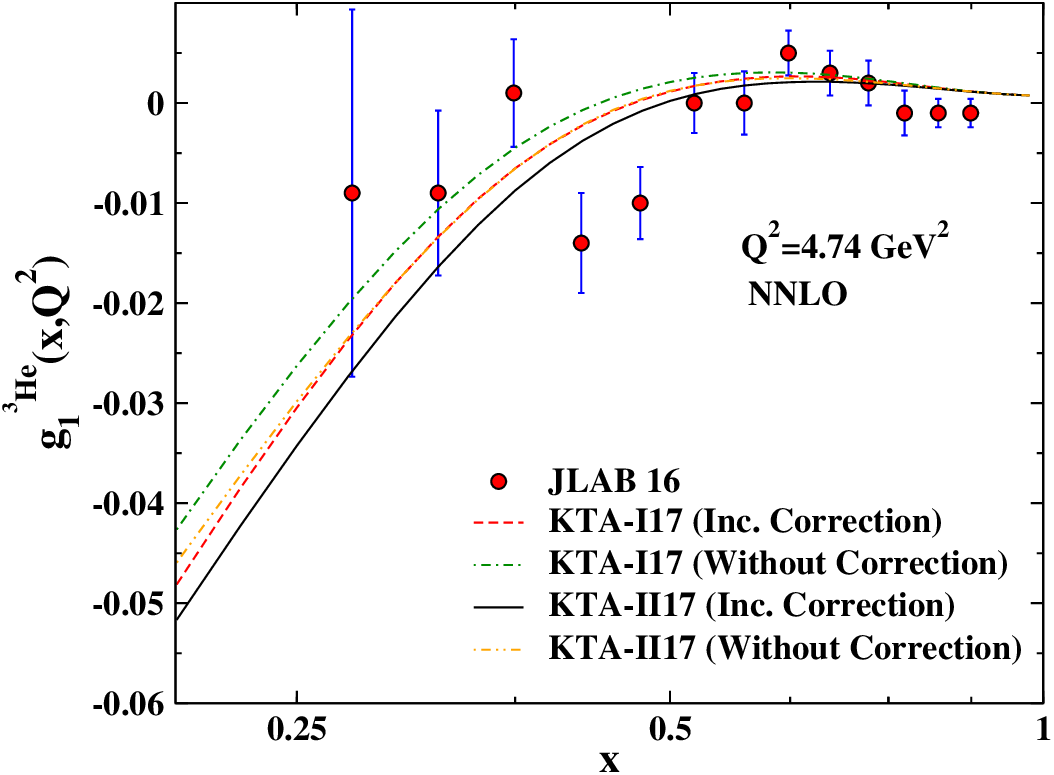}}   
		\resizebox{0.45\textwidth}{!}{\includegraphics{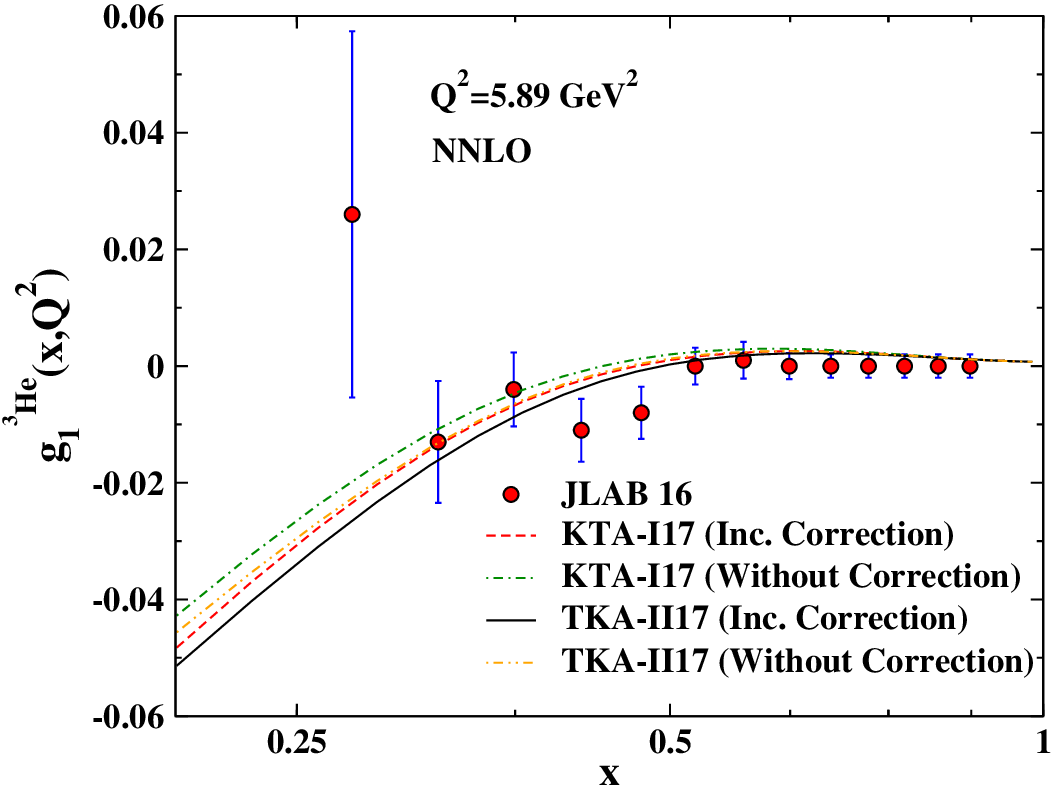}}   
		\caption{  Analytical result for the polarized $g_1^{^3{\mathrm He}} (x, Q^2)$ structure function as a function of $x$ at NNLO approximation that has been compared with the recent and up-to-date experimental data from JLAB16 Collaboration~\cite{Flay:2016wie}. The left plot corresponds to ${\rm Q}^2 = 4.74 \, {\rm GeV}^2$ and the right one corresponds to ${\rm Q}^2 = 5.89 \, {\rm GeV}^2$. }\label{fig:g1He3Q4p75-g1He3Q5p89}
	\end{center}
\end{figure*}

\begin{figure*}[htb]
	\begin{center}
		\vspace{0.50cm}
		\resizebox{0.45\textwidth}{!}{\includegraphics{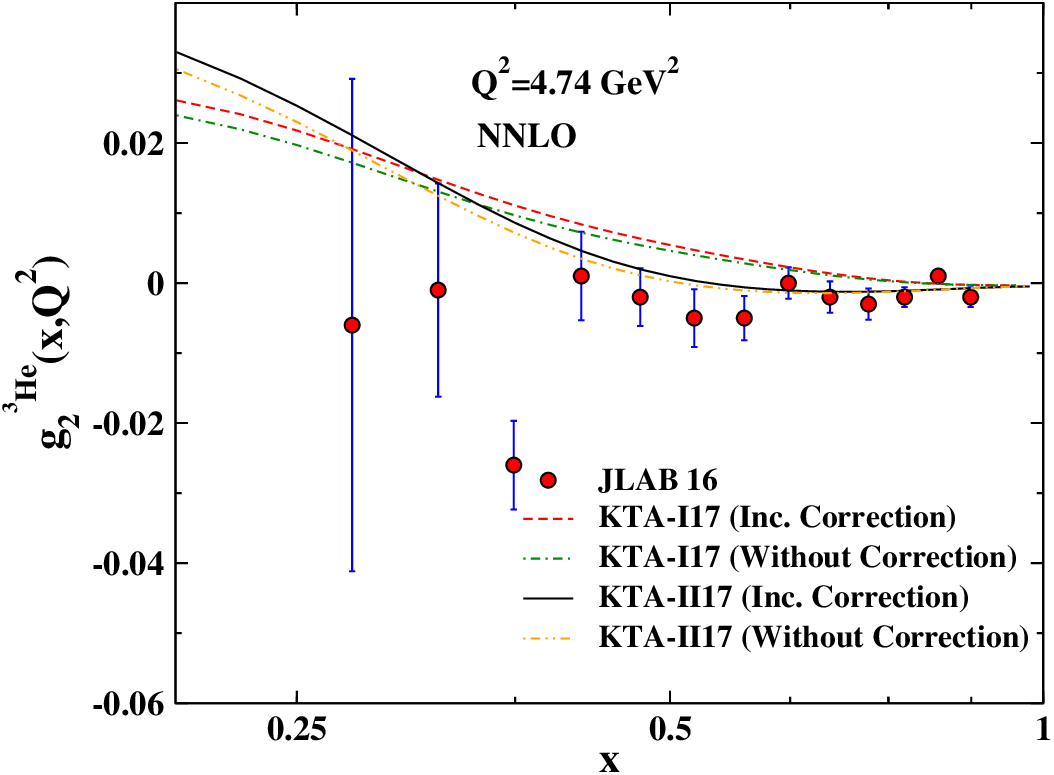}}   
		\resizebox{0.45\textwidth}{!}{\includegraphics{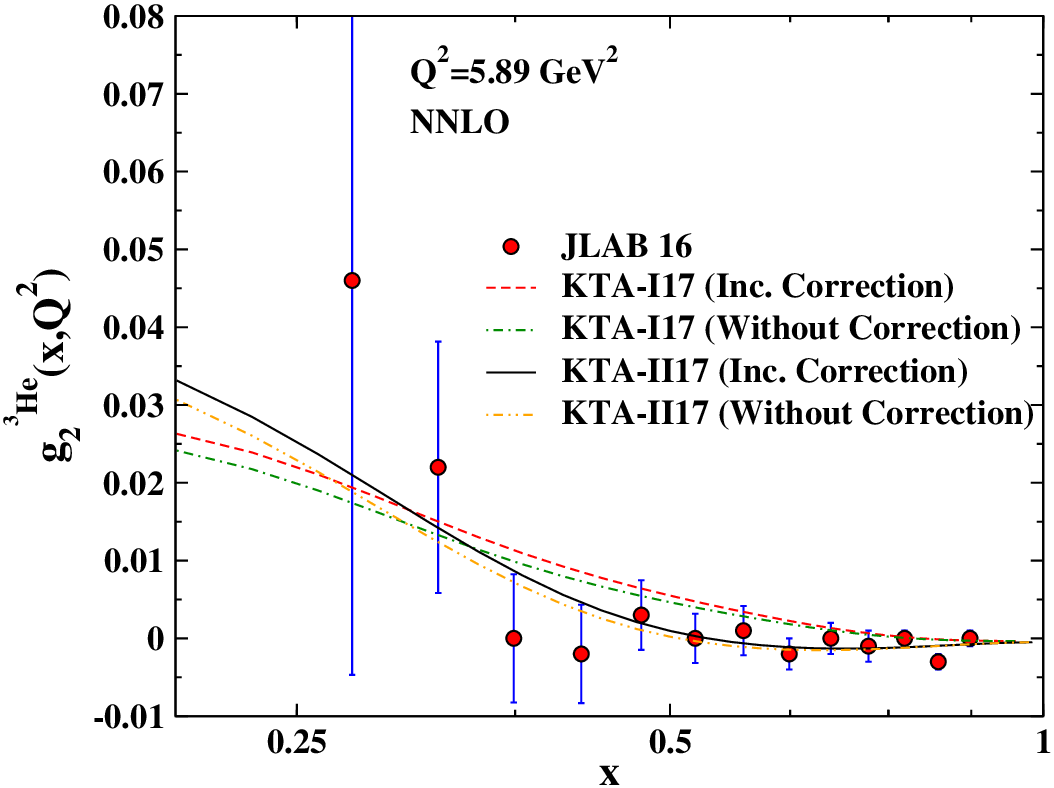}}   
		\caption{  Analytical result for the polarized $g_2^{^3{\mathrm He}} (x, Q^2)$ structure function as a function of $x$ at the NNLO approximation, which has been compared with the recent and up-to-date experimental data from the JLAB16 Collaboration~\cite{Flay:2016wie}. The left plot corresponds to ${\rm Q}^2 = 4.74 \, {\rm GeV}^2$ and the right one corresponds to ${\rm Q}^2 = 5.89 \, {\rm GeV}^2$.}\label{fig:g2He3Q4p74-g2He3Q5p89}
	\end{center}
\end{figure*}

In Figs.~\ref{fig:g1He3Q4p75-g1He3Q5p89} and \ref{fig:g2He3Q4p74-g2He3Q5p89}, our theory predictions for the polarized $g_1^{^3{\mathrm He}} (x, Q^2)$ and $g_2^{^3{\mathrm He}} (x, Q^2)$ structure functions are displayed as a function of $x$ at NNLO approximation and compared with the recent data from the JLAB16 Collaboration~\cite{Flay:2016wie}. The left plots correspond to ${\rm Q}^2 = 4.74 \, {\rm GeV}^2$ and the right ones correspond to ${\rm Q}^2 = 5.89 \, {\rm GeV}^2$.
From Fig.~\ref{fig:g1He3Q4p75-g1He3Q5p89}, we can conclude that applying nuclear corrections to the spin-dependent $^3{\mathrm He}$ structure functions decreases $g_1^{^3{\mathrm He}} (x, Q^2)$ at small values of $x$.
We observe similar trends for $g_2^{^3{\mathrm He}} (x, Q^2)$ in Fig.~\ref{fig:g2He3Q4p74-g2He3Q5p89}.
Figure~\ref{fig:g2He3Q4p74-g2He3Q5p89} also indicates that the results are strongly model dependent in the whole Bjorken $x$ region.
The $g_2^{^3{\mathrm He}}(x, Q^2)$ structure functions based on KTA-II17 polarized PDF describe the JLAB16 data better than KTA-I17 PPDFs.

The DIS data reported by E06-014 experiments at Jefferson Lab (JLAB) in Hall A are the latest and most up-to-date data for the spin-dependent $g_1$ and $g_2$ structure functions of $^3{\mathrm He}$~\cite{Flay:2016wie}.
These data sets were obtained from the scattering of a longitudinally polarized electron beam from a transversely and longitudinally polarized $^3{\mathrm He}$ target. This measurement covers the kinematic regions of $0.25 \leq x \leq 0.9$ and $2 \, {\rm GeV}^2 \leq {\rm Q}^2 \leq 6 \, {\rm GeV}^2$.
%
%
%

\begin{figure}[htb]
	\includegraphics[clip,width=0.450\textwidth]{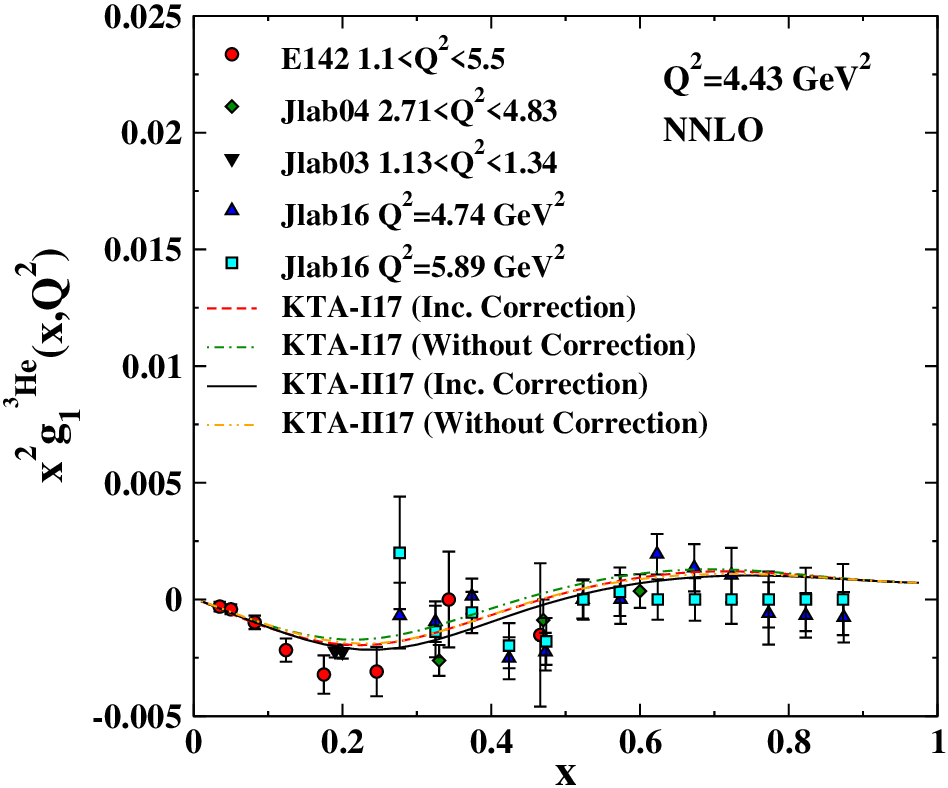}
	\begin{center}
		\caption{{\small  KTA-II17 results for the $x^2 g_1^{^3{\mathrm He}} (x, Q^2)$ structure function  as a function of $x$ at the NNLO approximation.
		Our curves are compared to the world data from E142~\cite{Anthony:1996mw}, JLAB04~\cite{Zheng:2004ce}, JLAB03~\cite{Kramer:2003un} and recent data from the JLAB16 Collaboration~\cite{Flay:2016wie}.  \label{fig:xxg1He3}}}
	\end{center}
\end{figure}

In Fig.~\ref{fig:xxg1He3}, the spin-dependent $x^2 g_1^{^3{\mathrm He}} (x, Q^2)$ structure function is plotted and compared to the world DIS data from E142~\cite{Anthony:1996mw}, JLAB04~\cite{Zheng:2004ce}, JLAB03~\cite{Kramer:2003un} and recent data from JLAB16 collaboration~\cite{Flay:2016wie}. KTA-II17 predictions follow the trend of existing data.
%

%
%
\subsection{Bjorken sum rule}\label{Bjorken-sum-rule}

Having at hand the spin-dependent structure of the proton and neutron, one may also examine in more detail the Bjorken sum rule~\cite{Bjorken:1966jh}
%
\begin{equation}\label{eq:Bjorken-sum}
\int_{o}^{1} [g_1^p(x, Q^2) - g_1^n(x, Q^2)] dx = \frac{1}{6} g_A [ 1 + {\cal O} (\frac{\alpha_s}{\pi}) ]\,,
\end{equation}
%
which relates the difference of the first moments of the proton $\int_0^1 g_1^p(x, Q^2) \, dx$ and neutron $\int_0^1 g_1^n(x, Q^2) \, dx$ spin structure functions to the axial vector
coupling constant measured in the $\beta$-decay of neutrons, $g_A = 1.2670 \pm 0.0035$~\cite{Olive:2016xmw}.
This sum rule can be straightforwardly generalized for the difference of the spin structure functions of $^3{\mathrm He}$ and $^3{\mathrm H}$ as follows~\cite{Bissey:2001cw,Boros:2000af}
%
\begin{equation}\label{eq:Bjorken-sum-He}
\int_{0}^{3} [g_1^{^3{\mathrm H}}(x,Q^2)-g_1^{^3{\mathrm He}}(x,Q^2)]dx = \frac{1}{6} g_A|_{triton} [ 1 + {\cal O} (\frac{\alpha_s}{\pi}) ]\,,
\end{equation}
%
where $g_A|_{triton}$ is the axial vector coupling constant measured in the $\beta$ decay of the triton, with $g_A|_{triton} = 1.211 \pm 0.002$~\cite{Budick:1991zb}.
Finally, taking the ratio of  Eqs.~(\ref{eq:Bjorken-sum}) and~(\ref{eq:Bjorken-sum-He}), one can find~\cite{Bissey:2001cw,Boros:2000af}
%
\begin{eqnarray}\label{eq:sumratio}
\eta &\equiv& \frac{g_A|_{triton}}{g_A} =  \nonumber \\
&& \frac{\int_0^3[g_1^{^3{\mathrm H}}(x,Q^2)-g_1^{^3{\mathrm He}}(x,Q^2)]dx} {\int_0^1[g_1^p(x,Q^2)-g_1^n(x,Q^2)]dx} = 0.9937 \pm 0.004 \,. \nonumber \\
\end{eqnarray}
%
We compute the above ratio to be $\eta = 0.923$ based on Eqs.~(\ref{g13He}) and~(\ref{g13H}).
Including nuclear corrections Eqs.~(\ref{g_1He-degrees-of-freedom}) and~(\ref{g_1H-degrees-of-freedom}) modifies this value to $\eta = 0.970$.

One can conclude that the corrections associated with the presence of the $\Delta$ resonance change the value of the Bjorken sum rule in the $A=3$ nuclei.
It has been shown in Refs.~\cite{Boros:2000af,Bissey:2001cw} that the contributions to the spin-dependent structure functions of $^3{\mathrm He}$ from non-nucleonic degrees of freedom in the nucleus, such as the $\Delta (1232)$ isobar, lead to the $\approx 4\%$ difference between the $g_A$ in the free nucleon and $g_A|_{triton}$ in the $A=3$ nuclei.

%
%
\section{Summary and conclusions} \label{Summary}

Our new NNLO analysis of the inclusive data included for the first time the extremely accurate COMPASS16 data on protons and deuterons~\cite{Adolph:2015saz,Adolph:2016myg}.
During the analysis, we considered TMCs to extract the polarized PDFs inside
the nucleon. We adopt more general input parametrizations for the sum of quark and antiquark polarized PDFs instead of the valence and sea quark distributions.
Similar to our previous papers, we use the Jacobi polynomial method to yield the structure functions $g_1^{\rm{N = p, n, d}}(x, Q^{2})$ from its moments in the whole $x$ region.
Having extracted the polarized structure functions, we estimated the nuclear structure functions of $g_{1,2}^{^3{\mathrm He}}$ and $g_{1}^{^3{\mathrm H}}$.

Due to increasing levels of precision attained in new generations of polarized DIS experiments, spin-dependent $^3{\mathrm He}$ and $^3{\mathrm H}$ targets become essential tools for studying the spin structure of the nucleons. They are also providing the most direct means of probing the polarized quarks and gluon distributions in the free neutron.

We also have performed a detailed analysis of nuclear corrections to the spin-dependent $g_{1,2}^{^3{\mathrm He}}$ and $g_{1}^{^3{\mathrm H}}$ structure functions. In addition to the nuclear effects arising from the ``incoherent'' scattering on nuclear targets, we have also examined the contributions from ``non-nucleonic degrees of freedom'' and have related the strength of these corrections to the Bjorken sum rule in the A=3 nuclei.

In this paper, we carry out an approximate way of assessing the importance of the nuclear effects. Neutron data are coming from deuterium and helium-3, and their corresponding nuclear corrections are ignored.
We believe that the neutron data are only reliable, if, we assume nuclear corrections are negligible whereas they are not. We suggest re-extracting the $g_1^n (x, Q^2)$ from $g_1^{^3{\mathrm He}} (x, Q^2)$ data as previously argued in Ref.~\cite{Bissey:2001cw}.

%
%
\section*{Acknowledgments}

The authors would like to thank Elliott Leader and Emanuele Nocera for their careful reading of the manuscript and helpful discussions.
The authors are also grateful to the School of Particles and Accelerators, Institute for Research in Fundamental Sciences (IPM) for financial support of this project.
Hamzeh Khanpour is grateful to the University of Science and Technology of Mazandaran for financial support provided for this research.

\appendix
%


%

\clearpage

\begin{thebibliography}{99}




\bibitem{Ball:2017nwa}
  R.~D.~Ball {\it et al.} [NNPDF Collaboration],
  ``Parton distributions from high-precision collider data,''
  \href{http://dx.doi.org/10.1140/epjc/s10052-017-5199-5}{{\rm  Eur.\ Phys.\ J.\ C} {\bfseries 77}, 663 (2017)}.




\bibitem{Bourrely:2015kla}
C.~Bourrely and J.~Soffer,
``New developments in the statistical approach of parton distributions: tests and predictions up to LHC energies,''
\href{http://dx.doi.org/10.1016/j.nuclphysa.2015.06.018}{{\rm Nucl.\ Phys.\ A} {\bfseries 941}, 307 (2015)}.


\bibitem{Harland-Lang:2014zoa}
L.~A.~Harland-Lang, A.~D.~Martin, P.~Motylinski and R.~S.~Thorne,
``Parton distributions in the LHC era: MMHT 2014 PDFs,''
\href{http://dx.doi.org/10.1140/epjc/s10052-015-3397-6}{{\rm Eur.\ Phys.\ J.\ C} {\bfseries 75}, no. 5, 204 (2015)}.






\bibitem{Martin:2009iq}
A.~D.~Martin, W.~J.~Stirling, R.~S.~Thorne and G.~Watt,
``Parton distributions for the LHC,''
\href{http://dx.doi.org/10.1140/epjc/s10052-009-1072-5}{{\rm Eur.\ Phys.\ J.\ C} {\bfseries 63},189 (2009)}.





\bibitem{Hou:2017khm}
T.~J.~Hou {\it et al.},
\href{https://arxiv.org/abs/1707.00657}{arXiv:1707.00657 [hep-ph].}






\bibitem{Shahri:2016uzl}
F.~Taghavi-Shahri, H.~Khanpour, S.~Atashbar Tehrani and Z.~Alizadeh Yazdi,
``Next-to-next-to-leading order QCD analysis of spin-dependent parton distribution functions and their uncertainties: Jacobi polynomials approach,''
\href{http://dx.doi.org/10.1103/PhysRevD.93.114024}{{\rm Phys.\ Rev.\ D} {\bfseries 93}, no. 11, 114024 (2016)}.





\bibitem{Jimenez-Delgado:2014xza}
P.~Jimenez-Delgado {\it et al.} [Jefferson Lab Angular Momentum (JAM) Collaboration],
``Constraints on spin-dependent parton distributions at large x from global QCD analysis,''
\href{http://dx.doi.org/10.1016/j.physletb.2014.09.049}{{\rm Phys.\ Lett.\ B} {\bfseries 738}, 263 (2014)}.




\bibitem{Sato:2016tuz}
N.~Sato {\it et al.} [Jefferson Lab Angular Momentum Collaboration],
``Iterative Monte Carlo analysis of spin-dependent parton distributions,''
\href{http://dx.doi.org/10.1103/PhysRevD.93.074005}{{\rm Phys.\ Rev.\ D} {\bfseries 93}, no. 7, 074005 (2016)}.




\bibitem{Leader:2014uua}
E.~Leader, A.~V.~Sidorov and D.~B.~Stamenov,
``New analysis concerning the strange quark polarization puzzle,''
\href{http://dx.doi.org/10.1103/PhysRevD.91.054017}{{\rm Phys.\ Rev.\ D} {\bfseries 91}, no. 5, 054017 (2015)}.




\bibitem{Nocera:2014gqa}
E.~R.~Nocera {\it et al.} [NNPDF Collaboration],
``A first unbiased global determination of polarized PDFs and their uncertainties,''
\href{http://dx.doi.org/10.1016/j.nuclphysb.2014.08.008}{{\rm Nucl.\ Phys.\ B} {\bfseries 887}, 276 (2014)}.





\bibitem{Khanpour:2016pph}
H.~Khanpour and S.~Atashbar Tehrani,
``Global Analysis of Nuclear Parton Distribution Functions and Their Uncertainties at Next-to-Next-to-Leading Order,''
\href{http://dx.doi.org/10.1103/PhysRevD.93.014026}{{\rm Phys.\ Rev.\ D} {\bfseries 93}, no. 1, 014026 (2016)}.




\bibitem{Eskola:2016oht}
K.~J.~Eskola, P.~Paakkinen, H.~Paukkunen and C.~A.~Salgado,
``EPPS16: Nuclear parton distributions with LHC data,''
\href{http://dx.doi.org/10.1140/epjc/s10052-017-4725-9}{{\rm Eur.\ Phys.\ J.\ C} {\bfseries 77}, no. 3, 163 (2017)}.



\bibitem{Kovarik:2015cma}
K.~Kovarik {\it et al.},
``nCTEQ15 - Global analysis of nuclear parton distributions with uncertainties in the CTEQ framework,''
\href{http://dx.doi.org/10.1103/PhysRevD.93.085037}{{\rm Phys.\ Rev.\ D} {\bfseries 93}, no. 8, 085037 (2016)}.



\bibitem{Wang:2016mzo}
R.~Wang, X.~Chen and Q.~Fu,
``Global study of nuclear modifications on parton distribution functions,''
\href{http://dx.doi.org/10.1016/j.nuclphysb.2017.04.008}{{\rm Nucl.\ Phys.\ B} {\bfseries 920}, 1 (2017)}.




\bibitem{Bertone:2017tyb}
  V.~Bertone {\it et al.} [NNPDF Collaboration],
  ``A determination of the fragmentation functions of pions, kaons, and protons with faithful uncertainties,''
  \href{http://dx.doi.org/10.1140/epjc/s10052-017-5088-y}{{\rm  Eur.\ Phys.\ J.\ C} {\bfseries 77}, 516 (2017)}.





\bibitem{Goharipour:2017rjl}
M.~Goharipour and H.~Mehraban,
``Predictions for the Isolated Prompt Photon Production at the LHC at $ \sqrt s= $ 13TeV,''
\href{http://dx.doi.org/10.1155/2017/3802381}{{\rm Adv.\ High Energy Phys.} {\bfseries 2017}, 3802381 (2017)}.






\bibitem{Dahiya:2016wjf}
H.~Dahiya and M.~Randhawa,
``Nucleon structure functions and longitudinal spin asymmetries in the chiral quark constituent model,''
\href{http://dx.doi.org/10.1103/PhysRevD.93.114030}{{\rm Phys.\ Rev.\ D} {\bfseries 93}, no. 11, 114030 (2016)}.



\bibitem{Jimenez-Delgado:2013boa}
P.~Jimenez-Delgado, A.~Accardi and W.~Melnitchouk,
``Impact of hadronic and nuclear corrections on global analysis of spin-dependent parton distributions,''
\href{http://dx.doi.org/10.1103/PhysRevD.89.034025}{{\rm Phys.\ Rev.\ D} {\bfseries 89}, no. 3, 034025 (2014)}.





\bibitem{Ball:2016spl}
R.~D.~Ball, E.~R.~Nocera and J.~Rojo,
``The asymptotic behaviour of parton distributions at small and large $x$,''
\href{http://dx.doi.org/10.1140/epjc/s10052-016-4240-4}{{\rm Eur.\ Phys.\ J.\ C} {\bfseries 76}, no. 7, 383 (2016)}.




\bibitem{Goharipour:2017uic}
M.~Goharipour and H.~Mehraban,
``Study of isolated prompt photon production in $ p $-Pb collisions for the ALICE kinematics,''
\href{http://dx.doi.org/10.1103/PhysRevD.95.054002}{{\rm Phys.\ Rev.\ D} {\bfseries 95}, no. 5, 054002 (2017)}.





\bibitem{Haider:2016zrk}
H.~Haider, F.~Zaidi, M.~Sajjad Athar, S.~K.~Singh and I.~Ruiz Simo,
``Nuclear medium effects in $F_{2A}^{EM}(x,Q^2)$ and $F_{2A}^{Weak}(x,Q^2)$ structure functions,''
\href{http://dx.doi.org/10.1016/j.nuclphysa.2016.06.006}{{\rm Nucl.\ Phys.\ A} {\bfseries 955}, 58 (2016)}.



\bibitem{Accardi:2016qay}
A.~Accardi, L.~T.~Brady, W.~Melnitchouk, J.~F.~Owens and N.~Sato,
``Constraints on large-$x$ parton distributions from new weak boson production and deep-inelastic scattering data,''
\href{http://dx.doi.org/10.1103/PhysRevD.93.114017}{{\rm Phys.\ Rev.\ D} {\bfseries 93}, no. 11, 114017 (2016)}.




\bibitem{Armesto:2015lrg}
N.~Armesto, H.~Paukkunen, J.~M.~Penín, C.~A.~Salgado and P.~Zurita,
``An analysis of the impact of LHC Run I proton?lead data on nuclear parton densities,''
\href{http://dx.doi.org/10.1140/epjc/s10052-016-4078-9}{{\rm Eur.\ Phys.\ J.\ C} {\bfseries 76}, no. 4, 218 (2016)}.





\bibitem{Frankfurt:2015cwa}
L.~Frankfurt, V.~Guzey, M.~Strikman and M.~Zhalov,
``Nuclear shadowing in photoproduction of ? mesons in ultraperipheral nucleus collisions at RHIC and the LHC,''
\href{http://dx.doi.org/10.1016/j.physletb.2015.11.012}{{\rm Phys.\ Lett.\ B} {\bfseries 752}, 51 (2016)}.




\bibitem{Frankfurt:2016qca}
L.~Frankfurt, V.~Guzey and M.~Strikman,
``Dynamical model of antishadowing of the nuclear gluon distribution,''
\href{http://dx.doi.org/10.1103/PhysRevC.95.055208}{{\rm Phys.\ Rev.\ C} {\bfseries 95}, no. 5, 055208 (2017)}.


\bibitem{MoosaviNejad:2016qdx}
S.~M.~Moosavi Nejad and P.~Sartipi Yarahmadi,
\href{http://dx.doi.org/10.1140/epja/i2016-16315-7}{{\rm Eur. Phys. J. A} {\bfseries 52}, 315 (2016)}.



\bibitem{Nejad:2015fdh}
S.~M.~Moosavi Nejad, M.~Soleymaninia and A.~Maktoubian,
\href{http://dx.doi.org/10.1140/epja/i2016-16316-6}{{\rm Eur. Phys. J. A} {\bfseries 52}, 316 (2016)}.




\bibitem{Guzey:2016qwo}
V.~Guzey, M.~Strikman and M.~Zhalov,
``Accessing transverse nucleon and gluon distributions in heavy nuclei using coherent vector meson photoproduction at high energies in ion ultraperipheral collisions,''
\href{http://dx.doi.org/10.1103/PhysRevC.95.025204}{{\rm Phys.\ Rev.\ C} {\bfseries 95}, no. 2, 025204 (2017)}.





\bibitem{Khanpour:2017slc}
H.~Khanpour, M.~Goharipour and V.~Guzey,
``Effects of next-to-leading order DGLAP evolution on generalized parton distributions of the proton and deeply virtual Compton scattering at high energy,''
\href{https://arxiv.org/abs/1708.05740}{arXiv:1708.05740 [hep-ph].}





\bibitem{Salajegheh:2015xoa}
M.~Salajegheh,
``Intrinsic strange distributions in the nucleon from the light-cone models,''
\href{http://dx.doi.org/10.1103/PhysRevD.92.074033}{{\rm Phys.\ Rev.\ D} {\bfseries 92}, no. 7, 074033 (2015)}.







\bibitem{Kalantarians:2017mkj}
  N.~Kalantarians, C.~Keppel and M.~E.~Christy,
  ``Comparison of the Structure Function F2 as Measured by Charged Lepton and Neutrino Scattering from Iron Targets,''
  \href{http://dx.doi.org/10.1103/PhysRevC.96.032201}{{\rm Phys.\ Rev.\ C} {\bfseries 96}, no. 3, 032201 (2017)}.





\bibitem{Ethier:2017zbq}
  J.~J.~Ethier, N.~Sato and W.~Melnitchouk,
  ``First simultaneous extraction of spin-dependent parton distributions and fragmentation functions from a global QCD analysis,''
  \href{http://dx.doi.org/10.1103/PhysRevLett.119.132001}{{\rm Phys.\ Rev.\ Lett.} {\bfseries 119}, no. 13, 132001 (2017)}.



\bibitem{Kusina:2016fxy}
  A.~Kusina {\it et al.},
  ``Vector boson production in pPb and PbPb collisions at the LHC and its impact on nCTEQ15 PDFs,''
   \href{http://dx.doi.org/10.1140/epjc/s10052-017-5036-x}{{\rm Eur.\ Phys.\ J.\ C} {\bfseries 77}, no. 7, 488 (2017)}.




\bibitem{Boroun:2015yea}
G.~R.~Boroun,
``Geometrical scaling behavior of the top structure functions ratio at the LHeC,''
\href{http://dx.doi.org/doi:10.1016/j.physletb.2015.03.051}{{\rm Phys.\ Lett.\ B} {\bfseries 744}, 142 (2015)}.



\bibitem{Boroun:2014nia}
G.~R.~Boroun,
``The ratio of the beauty structure functions $R^b = F^b_L/F^b_2$ at low $x$,''
\href{http://dx.doi.org/10.1016/j.nuclphysb.2014.05.010}{{\rm Nucl.\ Phys.\ B} {\bfseries 884}, 684 (2014)}.



\bibitem{Boroun:2014yea}
G.~R.~Boroun,
``Top structure function at the LHeC,''
\href{http://dx.doi.org/10.1016/j.physletb.2014.12.039}{{\rm Phys.\ Lett.\ B} {\bfseries 741}, 197 (2015)}.




\bibitem{Zarrin:2016kxf}
  S.~Zarrin and G.~R.~Boroun,
  ``Solution of QCD$\otimes$QED coupled DGLAP equations at NLO,''
  \href{http://dx.doi.org/10.1016/j.nuclphysb.2017.06.016}{{\rm  Nucl.\ Phys.\ B} {\bfseries 922}, 126 (2017)}.




\bibitem{AtashbarTehrani:2013qea}
S.~Atashbar Tehrani, F.~Taghavi-Shahri, A.~Mirjalili and M.~M.~Yazdanpanah,
``NLO analytical solutions to the polarized parton distributions, based on the Laplace transformation,''
\href{http://dx.doi.org/10.1103/PhysRevD.87.114012}{{\rm Phys.\ Rev.\ D} {\bfseries 87}, no. 11, 114012 (2013)},
\href{http://dx.doi.org/10.1103/PhysRevD.88.039902}{{\rm Erratum: Phys.\ Rev.\ D} {\bfseries 88}, no. 3, 039902 (2013)}.



\bibitem{TaghaviShahri:2010zz}
F.~Taghavi-Shahri and F.~Arash,
``Non-singlet spin structure function in the valon model and low x scaling behavior of $g_1^{NS}$ and $g_1^p$,''
\href{http://dx.doi.org/10.1103/PhysRevC.82.035205}{{\rm Phys.\ Rev.\ C} {\bfseries 82}, 035205 (2010)}.



\bibitem{Phukan:2017lzp}
  P.~Phukan, M.~Lalung and J.~K.~Sarma,
  ``NNLO solution of nonlinear GLR-MQ evolution equation to determine gluon distribution function using Regge like ansatz,''
  \href{http://dx.doi.org/10.1016/j.nuclphysa.2017.09.003}{{\rm  Nucl.\ Phys.\ A } {\bfseries 968}, 275 (2017)}.



\bibitem{Mottaghizadeh:2017vef}
  M.~Mottaghizadeh, F.~Taghavi Shahri and P.~Eslami,
  ``Analytical solutions of the QED$\otimes$QCD DGLAP evolution equations based on the Mellin transform technique,''
  \href{http://dx.doi.org/10.1016/j.physletb.2017.08.049}{{\rm  Phys.\ Lett.\ B } {\bfseries 773}, 375 (2017)}.




\bibitem{Ashman:1987hv}
J.~Ashman {\it et al.} [European Muon Collaboration],
``A Measurement of the Spin Asymmetry and Determination of the Structure Function $g_1$ in Deep Inelastic Muon-Proton Scattering,''
\href{http://dx.doi.org/10.1016/0370-2693(88)91523-7}{{\rm Phys.\ Lett.\ B} {\bfseries 206}, 364 (1988)}.



\bibitem{Ashman:1989ig}
J.~Ashman {\it et al.} [European Muon Collaboration],
``An Investigation of the Spin Structure of the Proton in Deep Inelastic Scattering of Polarized Muons on Polarized Protons,''
\href{http://dx.doi.org/10.1016/0550-3213(89)90089-8}{{\rm Nucl.\ Phys.\ B} {\bfseries 328}, 1 (1989)}.




\bibitem{Adeva:1998vv}
B.~Adeva {\it et al.} [Spin Muon Collaboration],
``{Spin asymmetries $A_1$ and structure functions $g_1$ of the proton and the deuteron from polarized high-energy muon scattering},''
\href{http://dx.doi.org/10.1103/PhysRevD.58.112001}{{\rm Phys. Rev.\ D} {\bfseries 58}, 112001 (1998)}.



\bibitem{Khanpour:2017cha}
H.~Khanpour, S.~T.~Monfared and S.~Atashbar Tehrani,
``Nucleon spin structure functions at NNLO in the presence of target mass corrections and higher twist effects,''
\href{http://dx.doi.org/10.1103/PhysRevD.95.074006}{{\rm Phys. Rev.\ D} {\bfseries 95}, no. 7, 074006 (2017)}.




\bibitem{Khanpour:2016uxh}
H.~Khanpour, A.~Mirjalili and S.~Atashbar Tehrani,
``Analytic derivation of the next-to-leading order proton structure function $F_2^p(x, Q^2)$ based on the Laplace transformation,''
\href{http://dx.doi.org/10.1103/PhysRevC.95.035201}{{\rm Phys. Rev.\ C} {\bfseries 95}, no. 3, 035201 (2017)}.



\bibitem{Ayala:2015epa}
C.~Ayala and S.~V.~Mikhailov,
``How to perform a QCD analysis of DIS in analytic perturbation theory,''
\href{http://dx.doi.org/10.1103/PhysRevD.92.014028}{{\rm Phys. Rev. D} {\bfseries 92}, no. 1, 014028 (2015)}.




\bibitem{Leader:1997kw}
E.~Leader, A.~V.~Sidorov and D.~B.~Stamenov,
``{NLO QCD analysis of polarized deep inelastic scattering},''
\href{http://dx.doi.org/10.1142/S0217751X98002547}{{\rm Int.\ J.\ Mod.\ Phys.\ A} {\bfseries 13}, 5573 (1998)}.






\bibitem{MoosaviNejad:2016ebo}
S.~M.~Moosavi Nejad, H.~Khanpour, S.~Atashbar Tehrani and M.~Mahdavi,
``QCD analysis of nucleon structure functions in deep-inelastic neutrino-nucleon scattering: Laplace transform and Jacobi polynomials approach,''
\href{http://dx.doi.org/10.1103/PhysRevC.94.045201}{{\rm Phys. Rev.\ C} {\bfseries 94}, no. 4, 045201 (2016)}.




\bibitem{Monfared:2014nta}
S.~Taheri Monfared, Z.~Haddadi and A.~N.~Khorramian,
``Target mass corrections and higher twist effects in polarized deep-inelastic scattering,''
\href{http://dx.doi.org/10.1103/PhysRevD.89.119901}{{\rm Phys. Rev.\ D} {\bfseries 89}, no. 7, 074052 (2014)}.
\href{http://dx.doi.org/10.1103/PhysRevD.89.074052}{{\rm Erratum: Phys. Rev.\ D} {\bfseries 89}, no. 11, 119901 (2014)}.





\bibitem{Leader:2002ni}
E.~Leader, A.~V.~Sidorov and D.~B.~Stamenov,
``On the role of higher twist in polarized deep inelastic scattering,''
\href{http://dx.doi.org/10.1103/PhysRevD.67.074017}{{\rm Phys. Rev. D} {\bfseries 67}, 074017 (2003)}.





\bibitem{Leader:2006xc}
E.~Leader, A.~V.~Sidorov and D.~B.~Stamenov,
``Impact of CLAS and COMPASS data on Polarized Parton Densities and Higher Twist,''
\href{http://dx.doi.org/10.1103/PhysRevD.75.074027}{{\rm Phys.\ Rev.\ D} {\bfseries 75},  074027 (2007)}.




\bibitem{Blumlein:2010rn}
J.~Blumlein and H.~Bottcher,
\href{http://dx.doi.org/10.1016/j.nuclphysb.2010.08.005}{{\rm Nucl.\ Phys.\ B} {\bfseries 841}, 205 (2010)}.




\bibitem{Adolph:2016myg}
C.~Adolph {\it et al.} [COMPASS Collaboration],
``Final COMPASS results on the deuteron spin-dependent structure function $g_1^{\rm d}$ and the Bjorken sum rule,''
\href{http://dx.doi.org/10.1016/j.physletb.2017.03.018}{{\rm Phys.\ Lett.\ B} {\bfseries 769},  34 (2017)}.



\bibitem{Adolph:2015saz}
C.~Adolph {\it et al.} [COMPASS Collaboration],
``{The spin structure function $g_1^{\rm p}$ of the proton and a test of the Bjorken sum rule},''
\href{http://dx.doi.org/10.1016/j.physletb.2015.11.064}{{\rm Phys.\ Lett.\ B} {\bfseries 753}, 18 (2016)}.





\bibitem{Abe:1998wq}
K.~Abe {\it et al.} [E143 Collaboration],
``{Measurements of the proton and deuteron spin structure functions $g_1$ and $g_2$}''
\href{http://dx.doi.org/10.1103/PhysRevD.58.112003}{{\rm Phys. Rev.\ D} {\bfseries 58}, 112003 (1998)}.





\bibitem{E155d}
P.~L.~Anthony \textit{et al.} {[}E155 Collaboration{]},
``{Measurement of the deuteron spin structure function $g_1^d(x)$ for
	$1(GeV/c)^2 < Q^2 < 40(GeV/c)^2$},''
\href{http://dx.doi.org/10.1016/S0370-2693(99)00940-5}{{\rm Phys.\ Lett.\ B} {\bfseries 463}, 339 (1999)}.




\bibitem{Ackerstaff:1997ws}
K.~Ackerstaff {\it et al.} [HERMES Collaboration],
``Measurement of the neutron spin structure function $g_1^n$ with a polarized $^3He$ internal target,''
\href{http://dx.doi.org/10.1016/S0370-2693(97)00611-4}{{\rm Phys.\ Lett.\ B} {\bfseries 404}, 383 (1997)}.





\bibitem{Abe:1997qk}
K.~Abe {\it et al.} [E154 Collaboration],
``Measurement of the neutron spin structure function g2(n) and asymmetry A2(n),''
\href{http://dx.doi.org/10.1016/S0370-2693(97)00613-8}{{\rm Phys.\ Lett.\ B} {\bfseries 404}, 377 (1997)}.




\bibitem{Abe:1997cx}
K.~Abe {\it et al.} [E154 Collaboration],
``Precision determination of the neutron spin structure function g1(n),''
\href{http://dx.doi.org/10.1103/PhysRevLett.79.26}{{\rm Phys.\ Rev.\ Lett.} {\bfseries 79}, 26 (1997)}.
[hep-ex/9705012].





\bibitem{Flay:2016wie}
D.~Flay {\it et al.} [Jefferson Lab Hall A Collaboration],
``Measurements of $d_{2}^{n}$ and $A_{1}^{n}$: Probing the neutron spin structure,''
\href{http://dx.doi.org/10.1103/PhysRevD.94.052003}{{\rm Phys. Rev.\ D} {\bfseries 94}, no. 5, 052003 (2016)}.





\bibitem{HERM98}
A.~Airapetian \textit{et al.} {[}HERMES Collaboration{]},   ``{Measurement of the proton spin structure function $g_1^p$ with a pure hydrogen target},''
\href{http://dx.doi.org/10.1016/S0370-2693(98)01341-0}{{\rm Phys.\ Lett.\ B} {\bfseries 442}, 484 (1998)}.




\bibitem{HERMpd}
A.~Airapetian \textit{et al.} {[}HERMES Collaboration{]},
`{`Precise determination of the spin structure function $g_1$ of the  proton,
	deuteron and neutron},''
\href{http://dx.doi.org/10.1103/PhysRevD.75.012007}{{\rm Phys. Rev.\ D} {\bfseries 75},  012007 (2007)}.



\bibitem{E155p}
P.~L.~Anthony \textit{et al.} {[}E155 Collaboration{]},
``{Measurements of the $Q^2$ dependence of the proton and neutron spin
	structure functions $g_1^p$ and $g_1^n$},''
\href{http://dx.doi.org/10.1016/S0370-2693(00)01014-5}{{\rm Phys.\ Lett.\ B} {\bfseries 493}, 19 (2000)}.



\bibitem{COMP1}
M.~G.~Alekseev \textit{et al.} {[}COMPASS Collaboration{]},
``{The Spin-dependent Structure Function of the Proton $g_1^p$ and a Test of the
	Bjorken Sum Rule},''
\href{http://dx.doi.org/10.1016/j.physletb.2010.05.069}{{\rm Phys.\ Lett.\ B} {\bfseries 690}, 466 (2010)}.
V.~Y.~Alexakhin \textit{et al.} {[}COMPASS Collaboration{]},
``{The Deuteron Spin-dependent Structure Function $g_1^d$ and its First Moment},''
\href{http://dx.doi.org/10.1016/j.physletb.2006.12.076}{{\rm Phys.\ Lett.\ B} {\bfseries 647}, 8 (2007)}.







\bibitem{E142n}
P.~L.~Anthony \textit{et al.} {[}E142 Collaboration{]},
``{Deep Inelastic Scattering of Polarized Electrons by Polarized $^3$He and
	the Study of the Neutron Spin Structure},''
\href{http://dx.doi.org/10.1103/PhysRevD.54.6620}{{\rm Phys. Rev.\ D} {\bfseries 54},  6620 (1996)}.





\bibitem{E154n}
K.~Abe \textit{et al.} {[}E154 Collaboration{]},
``{Precision determination of the neutron spin structure function $g_1^n$},''
\href{http://dx.doi.org/10.1103/PhysRevLett.79.26}{{\rm Phys.\ Rev.\ Lett.} {\bfseries 79},  26 (1997)}.




\bibitem{JLABn2003}
K.~M.~Kramer [Jefferson Lab E97-103 Collaboration],
``{The search for higher twist effects in the spin-structure functions of the neutron},''
\href{http://dx.doi.org/10.1063/1.1607208}{{\rm AIP Conf.\ Proc.} {\bfseries 675}, 615 (2003)}.







\bibitem{JLABn2004}
X.~Zheng {\it et al.}  [Jefferson Lab Hall A Collaboration],
``{Precision measurement of the neutron spin asymmetries and spin-dependent structure functions in the valence quark region},''
\href{http://dx.doi.org/10.1103/PhysRevC.70.065207}{{\rm Phys. Rev.\ C} {\bfseries 70},  065207 (2004)}.




\bibitem{JLABn2005}
K.~Kramer, D.~S.~Armstrong, T.~D.~Averett, W.~Bertozzi, S.~Binet, C.~Butuceanu, A.~Camsonne and G.~D.~Cates {\it et al.},
``{The $Q^2$-dependence of the neutron spin structure function $g^n_2$ at low $Q^2$},''
\href{http://dx.doi.org/10.1103/PhysRevLett.95.142002}{{\rm Phys.\ Rev.\ Lett.} {\bfseries 95},  142002 (2005)}.





\bibitem{COMP2005}
E.~S.~Ageev {\it et al.}  [COMPASS Collaboration],
``{Measurement of the spin structure of the deuteron in the DIS region},''
\href{http://dx.doi.org/10.1016/j.physletb.2005.03.025}{{\rm Phys.\ Lett.\ B} {\bfseries 612}, 154 (2005)}.




\bibitem{COMP2006}
V.~Y.~.Alexakhin {\it et al.}  [COMPASS Collaboration],
``{The Deuteron Spin-dependent Structure Function $g_1^d$ and its First Moment},''
\href{http://dx.doi.org/10.1016/j.physletb.2006.12.076}{{\rm Phys.\ Lett.\ B} {\bfseries 647}, 8 (2007)}.




\bibitem{Moch:2014sna}
S.~Moch, J.~A.~M.~Vermaseren and A.~Vogt,
``The Three-Loop Splitting Functions in QCD: The Helicity-Dependent Case,''
\href{http://dx.doi.org/10.1016/j.nuclphysb.2014.10.016}{{\rm Nucl.\ Phys.\ B} {\bfseries 889}, 351 (2014)}.




\bibitem{Lampe:1998eu}
  B.~Lampe and E.~Reya,
  ``Spin physics and polarized structure functions,''
   \href{http://dx.doi.org/10.1016/S0370-1573(99)00100-3}{{\rm Phys.\ Rept.} {\bfseries 332}, 1 (2000)}.




\bibitem{Zijlstra:1993sh}
  E.~B.~Zijlstra and W.~L.~van Neerven,
  ``Order $\alpha_s^2$ corrections to the polarized structure function $g_1 (x,Q^2)$,''
  \href{http://dx.doi.org/10.1016/0550-3213(94)90538-X}{{\rm Nucl.\ Phys.\ B} {\bfseries 417}, 61 (1994)}.
Erratum: [Nucl.\ Phys.\ B {\bf 426}, 245 (1994)]
Erratum: [Nucl.\ Phys.\ B {\bf 773}, 105 (2007)]
Erratum: [Nucl.\ Phys.\ B {\bf 501}, 599 (1997)].





\bibitem{Lacombe:1981eg}
M.~Lacombe, B.~Loiseau, R.~Vinh Mau, J.~Cote, P.~Pires and R.~de Tourreil,
``Parametrization of the deuteron wave function of the Paris n-n potential,''
\href{http://dx.doi.org/10.1016/0370-2693(81)90659-6}{{\rm Phys.\ Lett.\ B} {\bfseries 101}, 139 (1981)}.



\bibitem{Buck:1979ff}
W.~W.~Buck and F.~Gross,
``A Family of Relativistic Deuteron Wave Functions,''
\href{http://dx.doi.org/10.1103/PhysRevD.20.2361}{{\rm Phys. Rev.\ D} {\bfseries 20},  2361 (1979)}.




\bibitem{Zuilhof:1980ae}
M.~J.~Zuilhof and J.~A.~Tjon,
``Electromagnetic Properties of the Deuteron and the Bethe-Salpeter Equation with One Boson Exchange,''
\href{http://dx.doi.org/10.1103/PhysRevC.22.2369}{{\rm Phys. Rev.\ C} {\bfseries 22},  2369 (1980)}.




\bibitem{Wandzura:1977qf}
S.~Wandzura and F.~Wilczek,
``Sum Rules for Spin Dependent Electroproduction: Test of Relativistic Constituent Quarks,''
\href{http://dx.doi.org/10.1016/0370-2693(77)90700-6}{{\rm Phys.\ Lett.\ B} {\bfseries 72}, 195 (1977)}.




  \bibitem{Georgi:1976ve}
  H.~Georgi and H.~D.~Politzer,
  \href{http://dx.doi.org/10.1103/PhysRevD.14.1829}{{\rm Phys.\ Rev.\ D} {\bfseries 14}, 1829 (1976)}.



\bibitem{Blumlein:1998nv}
J.~Blumlein and A.~Tkabladze,
``Target mass corrections for polarized structure functions and new sum rules,''
\href{http://dx.doi.org/10.1016/S0550-3213(99)00289-8}{{\rm Nucl.\ Phys.\ B} {\bfseries 553}, 427 (1999)}.




\bibitem{Piccione:1997zh}
A.~Piccione and G.~Ridolfi,
``Target mass effects in polarized deep inelastic scattering,''
\href{http://dx.doi.org/10.1016/S0550-3213(97)00716-5}{{\rm Nucl.\ Phys.\ B} {\bfseries 513}, 301 (1998)}.




\bibitem{Dong:2006jm}
Y.~B.~Dong,
``Target mass corrections to proton spin structure functions and quark-hadron duality,''
\href{http://dx.doi.org/10.1016/j.physletb.2006.09.002}{{\rm Phys.\ Lett.\ B} {\bfseries 641}, 272 (2006)}.



\bibitem{Nachtmann:1973mr}
O.~Nachtmann,
``{Positivity constraints for anomalous dimensions},''
\href{http://dx.doi.org/10.1016/0550-3213(73)90144-2}{{\rm Nucl.\ Phys.\ B} {\bfseries 63}, 237 (1973)}.




\bibitem{Bass:2009ed}
S.~D.~Bass and A.~W.~Thomas,
``The Nucleon's octet axial-charge $g(A)^{(8)}$ with chiral corrections,''
\href{http://dx.doi.org/10.1016/j.physletb.2010.01.008}{{\rm Phys.\ Lett.\ B} {\bfseries 684}, 216 (2010)}.




\bibitem{Patrignani:2016xqp}
C.~Patrignani {\it et al.} [Particle Data Group],
``Review of Particle Physics,''
\href{http://dx.doi.org/10.1088/1674-1137/40/10/100001}{{\rm Chin.\ Phys.\ C} {\bfseries 40}, 100001 (2016)}.
Chin.\ Phys.\ C {\bf 40}, no. 10, 100001 (2016).



\bibitem{James:1994vla}
F.~James and M.~Roos,
``Minuit: A System For Function Minimization And Analysis Of The Parameter
Errors And Correlations,''
\href{http://dx.doi.org/10.1016/0010-4655(75)90039-9}{{\rm Comput.\ Phys.\ Commun.} {\bfseries 10}, 343 (1975)}.





  \bibitem{Pumplin:2001ct}
  J.~Pumplin, D.~Stump, R.~Brock, D.~Casey, J.~Huston, J.~Kalk, H.~L.~Lai and W.~K.~Tung,
  ``Uncertainties of predictions from parton distribution functions. 2. The Hessian method,''
  \href{http://dx.doi.org/10.1103/PhysRevD.65.014013}{{\rm Phys.\ Rev.\ D} {\bfseries 65}, 014013 (2001)}.


  \bibitem{Martin:2002aw}
  A.~D.~Martin, R.~G.~Roberts, W.~J.~Stirling and R.~S.~Thorne,
  ``Uncertainties of predictions from parton distributions. 1: Experimental errors,''
  \href{http://dx.doi.org/10.1140/epjc/s2003-01196-2}{{\rm Eur.\ Phys.\ J.\ C} {\bfseries 28}, 455 (2003)}.


  \bibitem{Hou:2016sho}
  T.~J.~Hou {\it et al.},
  ``Reconstruction of Monte Carlo replicas from Hessian parton distributions,''
  \href{https://arxiv.org/abs/1607.06066}{arXiv:1607.06066 [hep-ph]}.


\bibitem{Shoeibi:2017lrl}
S.~Shoeibi, H.~Khanpour, F.~Taghavi-Shahri and K.~Javidan,
``Determination of neutron fracture functions from a global QCD analysis of the leading neutron production at HERA,''
\href{http://dx.doi.org/10.1103/PhysRevD.95.074011}{{\rm Phys.\ Rev.\ D} {\bfseries 95}, no. 7, 074011 (2017)}.



\bibitem{Hirai:2003pm}
M.~Hirai {\it et al.} [Asymmetry Analysis Collaboration],
``Determination of polarized parton distribution functions and their uncertainties,''
\href{http://dx.doi.org/10.1103/PhysRevD.69.054021}{{\rm Phys.\ Rev.\ D} {\bfseries 69}, 054021 (2004)}.



\bibitem{Bissey:2001cw}
F.~R.~P.~Bissey, V.~A.~Guzey, M.~Strikman and A.~W.~Thomas,
``Complete analysis of spin structure function $g_1$ of $^3He$,''
\href{http://dx.doi.org/10.1103/PhysRevC.65.064317}{{\rm Phys.\ Rev.\ C} {\bfseries 65}, 064317 (2002)}.




\bibitem{Guzey:2000wh}
V.~Guzey,
``Nuclear shadowing in polarized DIS on $LiD_6$ at small x and its effect on the extraction of the deuteron spin structure function $g^d_1 (x,Q^2)$,''
 \href{http://dx.doi.org/10.1103/PhysRevC.64.045201}{{\rm Phys.\ Rev.\ C} {\bfseries 64}, 045201 (2001)}.




\bibitem{Frankfurt:1996nf}
L.~Frankfurt, V.~Guzey and M.~Strikman,
``The Nuclear effects in ($g_1^{He^3}$) and the Bjorken sum rule for A=3,''
\href{http://dx.doi.org/10.1016/0370-2693(96)00625-9}{{\rm Phys.\ Lett.\ B} {\bfseries 381}, 379 (1996)}.



\bibitem{Guzey:1999rq}
V.~Guzey and M.~Strikman,
``Nuclear effects in $g_1^A(x,Q^2)$ at small x in deep inelastic scattering on $^7Li$ and $^3He$,''
 \href{http://dx.doi.org/10.1103/PhysRevC.61.014002}{{\rm Phys.\ Rev.\ C} {\bfseries 61}, 014002 (2000)}.



\bibitem{Ethier:2013hna}
J.~J.~Ethier and W.~Melnitchouk,
``Comparative study of nuclear effects in polarized electron scattering from $^3He$,''
\href{http://dx.doi.org/10.1103/PhysRevC.88.054001}{{\rm Phys.\ Rev.\ C} {\bfseries 88}, no. 5, 054001 (2013)}.




\bibitem{Bissey:2000ed}
F.~R.~P.~Bissey, A.~W.~Thomas and I.~R.~Afnan,
``Structure functions for the three nucleon system,''
\href{http://dx.doi.org/10.1103/PhysRevC.64.024004}{{\rm Phys.\ Rev.\ C} {\bfseries 64}, 024004 (2001)}.



\bibitem{Afnan:2003vh}
I.~R.~Afnan, F.~R.~P.~Bissey, J.~Gomez, A.~T.~Katramatou, S.~Liuti, W.~Melnitchouk, G.~G.~Petratos and A.~W.~Thomas,
``Deep inelastic scattering from A = 3 nuclei and the neutron structure function,''
\href{http://dx.doi.org/10.1103/PhysRevC.68.035201}{{\rm Phys.\ Rev.\ C} {\bfseries 68}, 035201 (2003)}.




\bibitem{Piller:1999wx}
G.~Piller and W.~Weise,
``Nuclear deep inelastic lepton scattering and coherence phenomena,''
\href{http://dx.doi.org/10.1016/S0370-1573(99)00107-6}{{\rm Phys.\ Rept.} {\bfseries 330}, 1 (2000)}.



\bibitem{FernandezdeCordoba:1995pt}
P.~Fernandez de Cordoba, E.~Marco, H.~Muther, E.~Oset and A.~Faessler,
``Deep inelastic lepton scattering in nuclei at x > 1 and the nucleon spectral function,''
\href{http://dx.doi.org/10.1016/S0375-9474(96)00249-7}{{\rm Nucl.\ Phys.\ A} {\bfseries 611}, 514 (1996)}.
[nucl-th/9511038].



\bibitem{Kulagin:1994fz}
S.~A.~Kulagin, G.~Piller and W.~Weise,
``Shadowing, binding and off-shell effects in nuclear deep inelastic scattering,''
\href{http://dx.doi.org/10.1103/PhysRevC.50.1154}{{\rm Phys.\ Rev.\ C} {\bfseries 50}, 1154 (1994)}.



\bibitem{Kulagin:1994cj}
S.~A.~Kulagin, W.~Melnitchouk, G.~Piller and W.~Weise,
``Spin dependent nuclear structure functions: General approach with application to the deuteron,''
\href{http://dx.doi.org/10.1103/PhysRevC.52.932}{{\rm Phys.\ Rev.\ C} {\bfseries 52}, 932 (1995)}.




\bibitem{Kulagin:2007ph}
S.~A.~Kulagin and W.~Melnitchouk,
``Deuteron spin structure functions in the resonance and DIS regions,''
\href{http://dx.doi.org/10.1103/PhysRevC.77.015210}{{\rm Phys.\ Rev.\ C} {\bfseries 77}, 015210 (2008)}.




\bibitem{Kulagin:2008fm}
S.~A.~Kulagin and W.~Melnitchouk,
``Spin structure functions of $^3He$ at finite $Q^2$,''
\href{http://dx.doi.org/10.1103/PhysRevC.78.065203}{{\rm Phys.\ Rev.\ C} {\bfseries 78}, 065203 (2008)}.




\bibitem{Kulagin:2004ie}
S.~A.~Kulagin and R.~Petti,
``Global study of nuclear structure functions,''
\href{http://dx.doi.org/10.1016/j.nuclphysa.2005.10.011}{{\rm Nucl.\ Phys.\ A} {\bfseries 765}, 126 (2006)}.



\bibitem{CiofidegliAtti:1993zs}
C.~Ciofi degli Atti, S.~Scopetta, E.~Pace and G.~Salme,
``Nuclear effects in deep inelastic scattering of polarized electrons off polarized $^3He$ and the neutron spin structure functions,''
\href{http://dx.doi.org/10.1103/PhysRevC.48.R968}{{\rm Phys.\ Rev.\ C} {\bfseries 48}, R968 (1993)}.



\bibitem{Friar:1990vx}
J.~L.~Friar, B.~F.~Gibson, G.~L.~Payne, A.~M.~Bernstein and T.~E.~Chupp,
``Neutron polarization in polarized $^3He$ targets,''
\href{http://dx.doi.org/10.1103/PhysRevC.42.2310}{{\rm Phys.\ Rev.\ C} {\bfseries 42}, 2310 (1990)}.



\bibitem{Saito:1990aj}
T.~Y.~Saito, Y.~Wu, S.~Ishikawa and T.~Sasakawa,
``Triton beta decay,''
\href{http://dx.doi.org/10.1016/0370-2693(90)91586-Z}{{\rm Phys.\ Lett.\ B} {\bfseries 242}, 12 (1990)}.



\bibitem{Olive:2016xmw}
C.~Patrignani {\it et al.} [Particle Data Group],
``Review of Particle Physics,''
\href{http://dx.doi.org/10.1088/1674-1137/40/10/100001}{{\rm Chin.\ Phys.\ C} {\bfseries 40}, no. 10, 100001 (2016)}.



\bibitem{deFlorian:2005mw}
D.~de Florian, G.~A.~Navarro and R.~Sassot,
``Sea quark and gluon polarization in the nucleon at NLO accuracy,''
\href{http://dx.doi.org/10.1103/PhysRevD.71.094018}{{\rm Phys.\ Rev.\ D} {\bfseries 71}, 094018 (2005)}.




\bibitem{Anthony:1996mw}
P.~L.~Anthony {\it et al.} [E142 Collaboration],
``Deep inelastic scattering of polarized electrons by polarized $^3He$ and the study of the neutron spin structure,''
\href{http://dx.doi.org/10.1103/PhysRevD.54.6620}{{\rm Phys.\ Rev.\ D} {\bfseries 54}, 6620 (1996)}.



\bibitem{Zheng:2004ce}
X.~Zheng {\it et al.} [Jefferson Lab Hall A Collaboration],
``Precision measurement of the neutron spin asymmetries and spin-dependent structure functions in the valence quark region,''
\href{http://dx.doi.org/10.1103/PhysRevC.70.065207}{{\rm Phys.\ Rev.\ C} {\bfseries 70}, 065207 (2004)}.


\bibitem{Airapetian:2006vy}
A.~Airapetian {\it et al.} [HERMES Collaboration],
``Precise determination of the spin structure function $g_1$ of the proton, deuteron and neutron,''
\href{http://dx.doi.org/10.1103/PhysRevD.75.012007}{{\rm Phys.\ Rev.\ D} {\bfseries 75}, 012007 (2007)}.


\bibitem{Kramer:2003un}
K. Kramer, Ph.D. thesis, College of William and Mary,2003.



\bibitem{Bjorken:1966jh}
J.~D.~Bjorken,
``Applications of the Chiral $U(6) \bigotimes U(6)$ Algebra of Current Densities,''
\href{http://dx.doi.org/10.1103/PhysRev.148.1467}{{\rm Phys.\ Rev.} {\bfseries 148}, 1467 (1966)}.




\bibitem{Boros:2000af}
C.~Boros, V.~A.~Guzey, M.~Strikman and A.~W.~Thomas,
``Role of the Delta (1232) in DIS on polarized $^3He$ and extraction of the neutron spin structure function $g_1^n(x,Q^2)$,''
\href{http://dx.doi.org/10.1103/PhysRevD.64.014025}{{\rm Phys.\ Rev.\ D} {\bfseries 64}, 014025 (2001)}.








\end{thebibliography}
%

\end{document}